\documentclass[preprint,12pt]{elsarticle}

\usepackage{amssymb}

\usepackage{amsmath}
\usepackage{amssymb}
\usepackage{paralist}
\usepackage[font=small,labelfont=bf]{caption}
\usepackage{multirow}
\usepackage{xspace}
\usepackage{color}
\usepackage{xcolor}
\usepackage{ifthen}
\usepackage{url}
\usepackage{fancybox}
\usepackage{enumitem}
\usepackage{listings,algorithmic}
\usepackage{balance}
\usepackage{ifthen}

  \usepackage[tight,footnotesize]{subfigure}
  \usepackage{graphicx}
   \graphicspath{{pics/}}
   \DeclareGraphicsExtensions{.pdf,.jpeg,.png}

\newcommand{\eg}{e.g.,}
\newcommand{\ie}{i.e.,}

\newboolean{showcomments}
\setboolean{showcomments}{true} 
\ifthenelse{\boolean{showcomments}}
{\newcommand{\nbnote}[2]{
  \fcolorbox{blue}{yellow}{\bfseries\sffamily\scriptsize#1}
  {\sf\small\textit{#2}}
 }
}
{\newcommand{\nbnote}[2]{}
 
}

\newtheorem{mydef}{\textbf{Definition}}
\newcommand{\use}{depend}
\newcommand{\couse}{co-dependency}
\newcommand{\Couse}{Co-Dependency}
\newcommand{\diffusion}{adoption-diffusion}

\newcommand{\SUG}{SUG}
\newcommand{\pSUG}{P-SUG}
\newcommand{\useBy}{rev\text{-}depend}
\newcommand{\Family}{project}
\newcommand{\family}{Project}
\newcommand{\variety}{variety}
\newcommand{\pop}{popularity}

\newcommand{\RqOne}{\emph{Are we able to apply the \SUG~to a real-world super repository? and if so, do our metrics and visualizations provide useful recommendations?}\xspace}

\newcommand{\RqTwo}{\emph{Can we use the \SUG~model to describe dependencies between super repository types?}\xspace}

\newenvironment{hassanbox}%
{\begin{center}\vspace{0mm}\noindent\begin{Sbox}\begin{minipage}{0.95\columnwidth}}%
{\end{minipage}\end{Sbox}\fbox{\TheSbox}\end{center}\vspace{0mm}}

\journal{The Journal of Systems and Software}

\begin{document}

\begin{frontmatter}



\title{Modeling Library Dependencies and Updates in Large Software Repository Universes}

 \author[label1]{Raula Gaikovina Kula}
  \author[label2]{Coen De Roover}
  \author[label3]{Daniel M. German}
  \author[label1]{Takashi Ishio}
  \author[label1]{Katsuro Inoue}
 \address[label1]{Osaka University, Japan}
 \address[label2]{Vrije Universiteit Brussel, Brussels, Belgium}
 \address[label3]{University of Victoria, Canada}


\begin{abstract}
Popular (re)use of third-party open-source software (OSS) is evidence of the impact of hosting repositories like maven on software development today. Updating libraries is crucial, with recent studies highlighting the associated vulnerabilities with aging OSS libraries. The decision to migrate to a newer library can range from trivial (security threat) to complex (assessment of work required to accommodate the changes).  By leveraging the `wisdom of the software repository crowd' we propose a simple and efficient approach to recommending `consented' library updates. Our Software Universe Graph (\SUG) models library dependency and update information mined from super repositories to provide different metrics and visualizations that aid in the update decision. To evaluate, we first constructed a \SUG~from 188,951 nodes of 6,374 maven unique artifacts. Then, we demonstrate how our metrics and visualizations are applied through real-world examples. As an extension, we show how the \SUG~can compare dependencies between different super repositories.  From a sample of 100 github applications, our method found that on average 79\% similar overlapping dependencies combinations exist between the maven and github super repository universes.
\end{abstract}


\end{frontmatter}



\section{Introduction}
The (re)use of third-party software  is now commonplace in today's software development, both open source software (OSS) and commercial settings alike~\cite{EbertOSS}, \cite{HainemannICSR2011}. Software libraries come with the promise of being able to reuse quality implementations, preventing \textit{`reinventions of the wheel'} and speeding up development. Examples of popular reuse libraries are the \textsc{Spring} \cite{springURL} web framework and the \textsc{Apache commons} \cite{commonsURL} collection of utility functions. Widespread use of OSS libraries has lead to massive stores of project repositories such as The Central Repository (Maven) \cite{MavenCentralURL}, Sourceforge \cite{sourceforgeURL} and Github \cite{githubURL}.  For instance, as of 05-10-2015,  maven central (\url{https://search.maven.org/#stats}) hosted over 120,000 unique projects.

Software is constantly evolving. With new versions continuously released, the maintenance of system's dependencies is not practiced enough. A study by Grinter identified aging libraries a threat to software livelihood \cite{Grinter}. In 2014, Sonatype reported that on average 24\% of buggy code in applications were linked to severe flaws in their outdated libraries.  That same year, the threat of high profile vulnerabilities Shellshock\footnote{\url{https://shellshocker.net/}}, HeartBleed\footnote{\url{http://heartbleed.com/}} and Poodle\footnote{\url{https://poodlebleed.com/}} highlighted the need to update dependencies in applications (also referred to as systems in this paper). Security vulnerabilities updates are a trivial decision as its threat to software quality outweighs the costs. Security experts recommend to update, regardless of the size of the changes to be made. 

More complex decisions are encountered when assessing the different risks and effort required to accommodate the changes. 
Studies \cite{mattsson1999framework}, \cite{Fayad:1999}, have reported that unless the underlying need is apparent, most maintainers are unmotivated or hesitant to update. Our previous work \cite{Kula2015} considered that developers exhibit a latency to migrate to the latest version released.

To this end, tools and techniques have been developed to address certain risks of migration. Take for instance, library incompatibility. Research tools such as \texttt{SemDiff} \cite{Dagenais:2009} and industry counterparts like \texttt{clirr} \cite{clirrrURL} are used to assist with library compatibility issues during migration. Moreover, other external technical, organizational or social factors also influence a maintainers decision to update. For instance, a maintainers personal preference or compliance to the organizational practices may influence the decision. These techniques though effective, only solve a specific risk.



With the advancements in online repository usage and data mining, we provide a much more efficient and simpler solution to library update recommendations. Building on our previous work on visualizing the evolution of a system and its library dependencies~\cite{2014VISSOFTKula} and on popular dependency combinations~\cite{YanoICPC2015},
we introduce the Software Universe Graph (SUG) as a generic means to  quantify and visualize ``wisdom-of-the-crowd'' insights for a software repository universe. We extend on the simple usage popularity metric with metrics to describe \emph{\diffusion} and \emph{\couse{}}. Our popularity is a measure of usage at any point in time.  The \textit{\diffusion} metrics measure the spread of library versions over all systems and the \textit{\couse}~metrics to describe how often two evolve library dependencies over time. For the evaluation, we show through real-world examples the application the \SUG~properties. The paper makes the following contributions:

\begin{itemize}

\item We introduce the graph-based \SUG~model to represent library dependency and update relationships within a large-scale super repository universe. We demonstrate practicality by construction from maven.

\item We extent on simple usage popularity to measure diffusion and co-dependency of libraries. The resulting recommendations are: 1.) visual prediction of either popular or obsolete software versions 2.) recommendation to adopt new library based on co-dependency in other systems and 3.) visualization of co-dependency evolution patterns between two library releases. 



\item We leverage the \SUG~co-dependency metric to compare dependency usage between super repository universes. We found Maven and Github systems to be overlapping, with on average 79\% similarity of dependencies. 
\end{itemize}

The paper layout is as follows. Section 2 details the motivation of the \SUG. Section 3 explains in detail the formal aspects of the \SUG~model. Section 4 introduced the metrics applied to the \SUG~model. Section 5 discusses the evaluation with the results presented in section 6. Discussions and related work are later shown in Section 7 and 8 respectability. Finally, we close with conclusions in Section 9.

\section{Mining the `wisdom of the crowd' from Super Repositories}
\label{sec:motivate}
Our approach involves studying the different library dependency relationships that exist in the super repository over time. Concretely, we are concerned with two aspects 1.) Diffusion of newer libraries and 2.) co-dependency patterns. 


According to the Diffusion of Innovations (DoI) theory~\cite{DoI}, 
successful technologies have different types of users: 
innovators, early adopters, the early majority, the late majority, and laggards. Applied to the super repository dependency relationships, we would like to understand the diffusion in terms of popular migration toward the different versions of libraries. Our rational is that crowd 'consent' of a library is evident by its successful adoption and diffusion over its predecessors. Our adoption-diffusion concept is inspired by use-diffusion~\cite{Venkatesh2004} metrics used in the field of economics and marketing.

The changes in the complex web of dependency relationships in the super repository characterizes the ripping effect of updating a single library dependency. The colloquial term `dependency hell', to describe these complexity of managing these dependencies. Maven and Gradle\footnote{\url{http://gradle.org/}} are examples of dependency management build tools employed for applications. We define these libraries used together in an application as co-dependencies. We conjecture that useful co-dependency patterns of the crowd can be inferred and used to recommend the best update combination for a set of libraries.

In this paper, we formulate a model in which both diffusion and co-dependency relationships can be captured, quantified and visualized using defined metrics of popularity, adoption-diffusion and co-dependency. Using a graph-based approach, we model dependency and update relations to handle all software systems in a super repository.

\section{The Super Repository Universe}

\begin{figure*}
\centering
\includegraphics[width=.9\columnwidth,clip, trim={0cm 0cm 0cm 0cm}]
{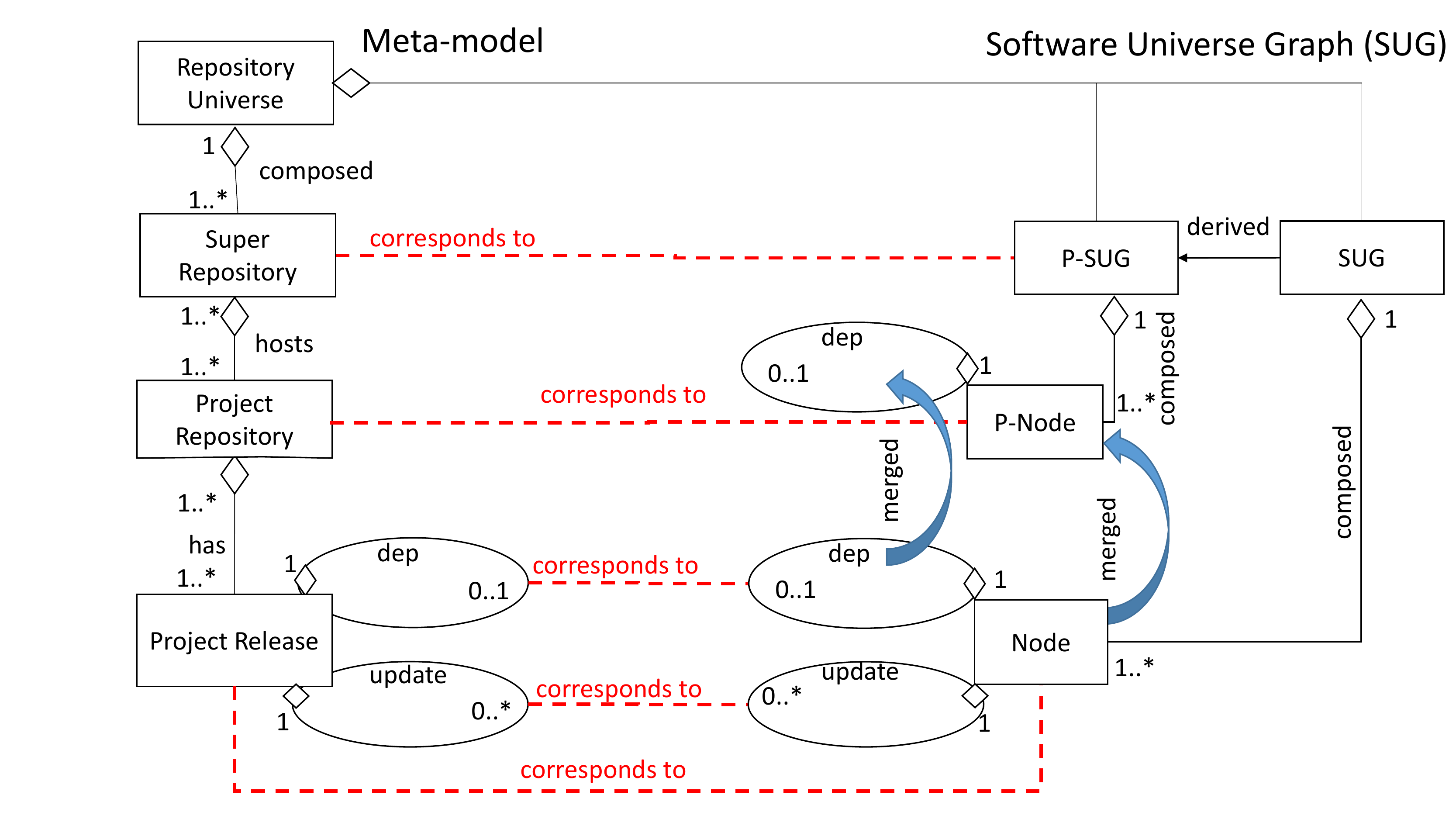}
\caption{We illustrate the linkage from the meta-model of the real world to the proposed \SUG~model.}
\label{fig:metalmodel}
\end{figure*}

\subsection{Modeling Super Software Repositories}
In this section, we show in Figure \ref{fig:metalmodel}  how our model handles the realities of library dependencies and update across software repositories.
We consider the virtual repository universe that encompasses both publicly accessible and private repositories.
We define a \textit{project release} as a published software unit with a version identifier.
For instance, version \texttt{3.6.3} of \texttt{SymmetricDs} ($SymmetricDs_{3.6.3}$).
A project release is either in source or in executable format.
 Examples of language-specific source code are \texttt{*.java, *.cpp, *.jss}  accompanied  by configuration build files. 
Executables are compiled binaries such as  \texttt{jar, exe or dll} files ready for (re)use. 
A project release may be superseded by  a newer project release, creating an update relationship.
Project releases can use other project releases as libraries and vice-versa, forming a dependency relation. 
Project releases linked by update relationships are managed by a \textit{project repository}. 
Project repositories may manage project release relations through project-specific conventions such as Semantic Versioning (SemVer)\footnote{\url{http://semver.org}}. 

The super repository hosts multiple project repositories. Related work refers to these as `super' repositories or repositories of repositories \cite{SulaymanRoR07}, \cite{LunguWCRE07}. We discern two types of super repositories: 
those that host libraries and those that host systems. Examples of library-hosting super repositories include \textsc{Maven} for JVM libraries, \textsc{RubyGems}\footnote{\url{https://rubygems.org}} for Ruby libraries, and \textsc{nuget}\footnote{\url{https://www.nuget.org}} for .NET and \textsc{npm}\footnote{\url{https://www.npmjs.com}} for JavaScript libraries. Examples of  system-hosting super repositories include \textsc{GitHub} and \textsc{Sourceforge} which primarily serve as hubs for collaborative development and end-user download respectively.


As depicted in Fig \ref{fig:metalmodel},  the \SUG~is an abstract representation of the realities of super repositories. Related studies  reveal web-like complex dependencies between project releases,  making the distinction between systems and libraries dependent on perspective \cite{2013CSMRGerman}, \cite{2013:wea:haenni}. Dependencies can even span across super repositories. Therefore, the model should not be restrictive in system or library identification. The model should also not be restrictive in implementation issues such as programming language and control version systems. Specifically there are two types of software universe models, the normal \SUG~(introduced in Section \ref{sec:SUG})~that corresponds to the project releases and the \pSUG~that correponds to dependencies at the project repository level. The \pSUG~is an aggregation of nodes and edges (merged dependency edges and dropped update edges) related to one particular project repository (later introduced in Section \ref{sec:pSUG}).

\subsection{The Software Universe Graph (\SUG)}
\begin{figure*}
\centering
\includegraphics[width=1.1\textwidth]{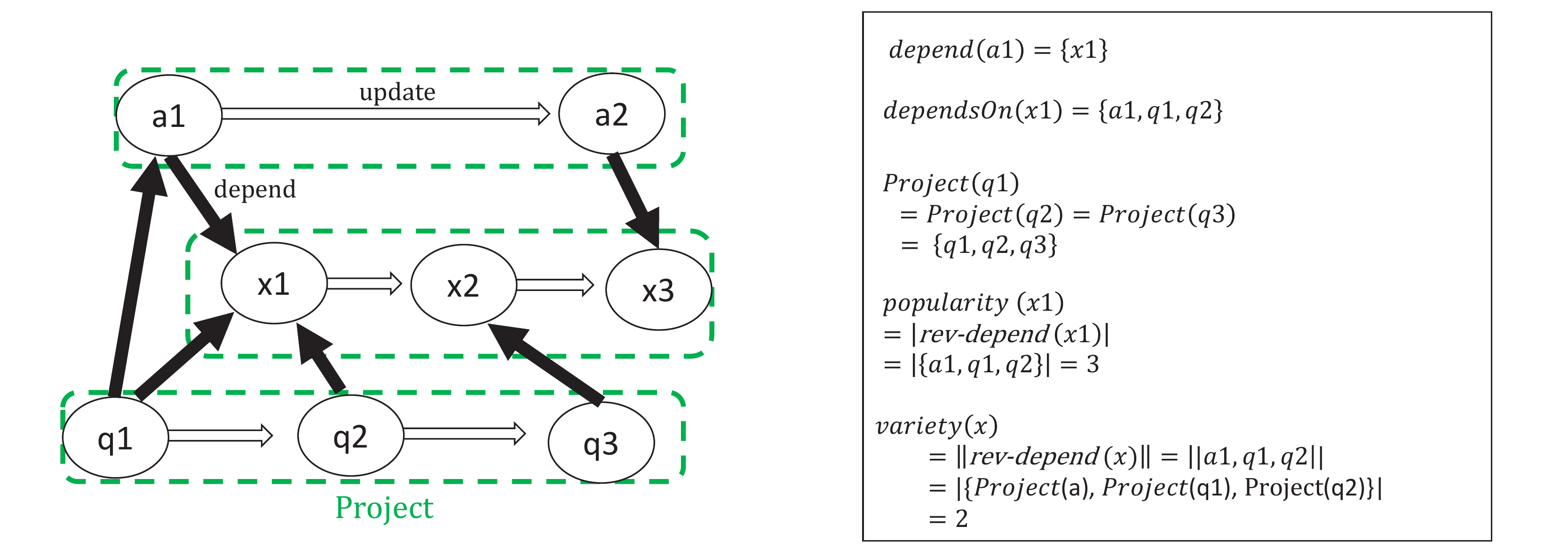}
\caption{Example of the Software Universe Graph}. \label{fig:SUG}
\end{figure*}

Figure \ref{fig:SUG} depicts the basic elements of the  Software Universe Graph. Let $G(N,E)$ be the SUG. $N$ is a set of nodes, with each node representing a project release instance.  For instance, \texttt{SymmetricDs} version \texttt{3.6.3} ($SymmetricDs_{3.6.3}$) is a project release instance represented as a single node. 

\label{sec:SUG}
For any \SUG, the edges $E$  are composed of $E_{dep}$ and $E_{up}$. $E_{dep}$ is a set of\textit{ dependency} edges and $E_{up}$ is a set of \textit{update} edges. 

\begin{mydef}
An edge $u \rightarrow v \in E_{dep}$ means that $u$ depends on  $v$ (\use). Reverse-dependency (\useBy) refers to the inverse. 
\end{mydef}

\begin{equation}
\use(u)\equiv \{v|u \rightarrow v\}
\end{equation}
\begin{equation}
\useBy(u)\equiv \{v|v \rightarrow u\}
\end{equation}

Dependency-relations can be extracted from either the source code or from build configuration files. As depicted in Figure \ref{fig:SUG}, node $a1$ (system) has a \use~relation to node $x1$ (library). Note that node $x1$ has reverse dependencies (\useBy)~to nodes $a1$, $q1$ and $q2$. Parallel edges for node pairs are not allowed. In this paper, we focus on popular project releases that are connected by many depend-relation edges.

\begin{mydef}
\textit{For a given node u, popularity is the number of incoming depend-relation edges.}
\end{mydef}

\begin{equation}
\pop(u)\equiv |\useBy(u)|
\end{equation}
For instance in Figure \ref{fig:SUG}, for node $x1$, $\pop(x1)=|\useBy(x1)|=|\{a1,q1,q2\}|=3 $.

\begin{mydef}
\textit{An edge $a \Rightarrow b \in E_{up}$ represents an update-relation from node $a$ to $b$, meaning $b$ is the immediate successor release of $a$.}
\end{mydef}

Update-relations refer to when a succeeding release of a project release is made available. Figure \ref{fig:SUG} shows that node $q1$ is first updated to node $q2$. Later on, node $q2$ is updated to the latest node $q3$. Hence, $q1 \Rightarrow q2 \Rightarrow q3$.

Let any SUG node $u$ be denoted by three attributes: \texttt{<name, release, time>}. For a node $u$, we define:

\begin{itemize}
\item \textbf{u.name} Name is the string representing the identifier of a software project.

For nodes $x$ and $y$, if $x \Rightarrow y$, then $x.name = y.name$ holds in the \SUG.

\item \textbf{u.release}. Release denotes the assigned change reference for a software project.
 For nodes $u$ and $v$, if $u \Rightarrow v$
then $v$ is the immediate successor of $u$. 

\item \textbf{u.time}. Time refers to the time-stamp at which node $u$ was released. For nodes $x$ and $y$ of $x \Rightarrow y$, $x.time < y.time$.
\end{itemize}

The \SUG{} node for the latest\footnote{\url{http://mvnrepository.com/artifact/junit/junit/4.11}: accessed 2014-08-02}  release of \textsc{junit}, for instance, is \texttt{<name = "junit", version= "4.11", time="2012-11-14">}.

\begin{figure}
\centering
\includegraphics[width=1.3\textwidth]{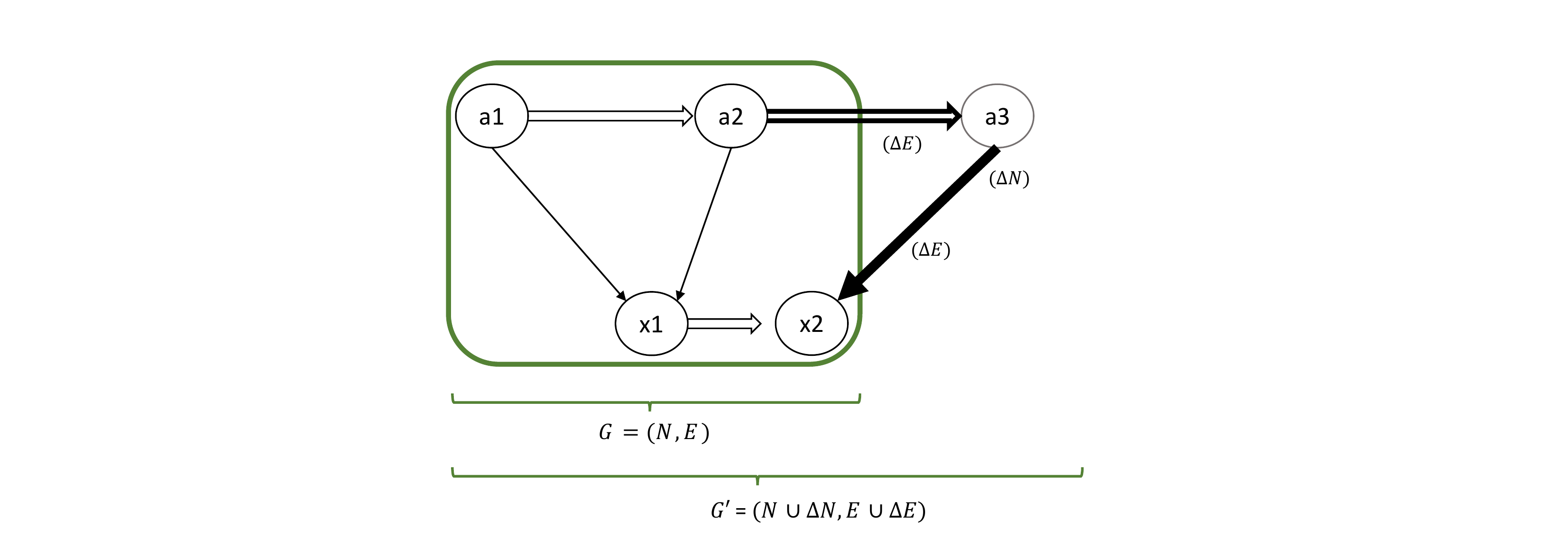}
\caption{Temporal property of the \SUG~}
\label{fig:SUGTemp}
\end{figure}

\begin{mydef}
\textit{A timed \SUG~specifies the state of the \SUG~at a given point in time.}
\end{mydef}

The temporal properties describe the simultaneity or ordering in reference to time. Let \SUG~$G = (N, E) $ be at time $t$. At time $t^{\prime} > t$, we observe an extension of $G$, such that: $G^{\prime} = (N \cup \Delta N, E \cup \Delta E)$ where $\Delta E \cap (N \times N) = \varnothing$. Figure \ref{fig:SUGTemp} depicts $G'$ composed of $G$ augmented with newly added node $a3$ and its corresponding $a3 \rightarrow x2$ and $a2 \Rightarrow a3$ relations. \SUG~$G_{t}$ at time $t$ is therefore a sub-graph of $G$. $Popularity_t(x)$ for a node $x$ at time $t$ can be described\footnote{ We define that $Popularity_t(x)=0 $ if $t< x.time$}. 

\subsection{The Project-level Software Universe Graph (P-SUG) }
\label{sec:pSUG}

Derived from the \SUG, the \pSUG~describes a set of nodes weakly connected by update-relations by \textit{\Family}. This corresponds to the project repository level described in Figure \ref{fig:metalmodel}. 

\begin{figure*}
\centering
\includegraphics[width=1\textwidth]{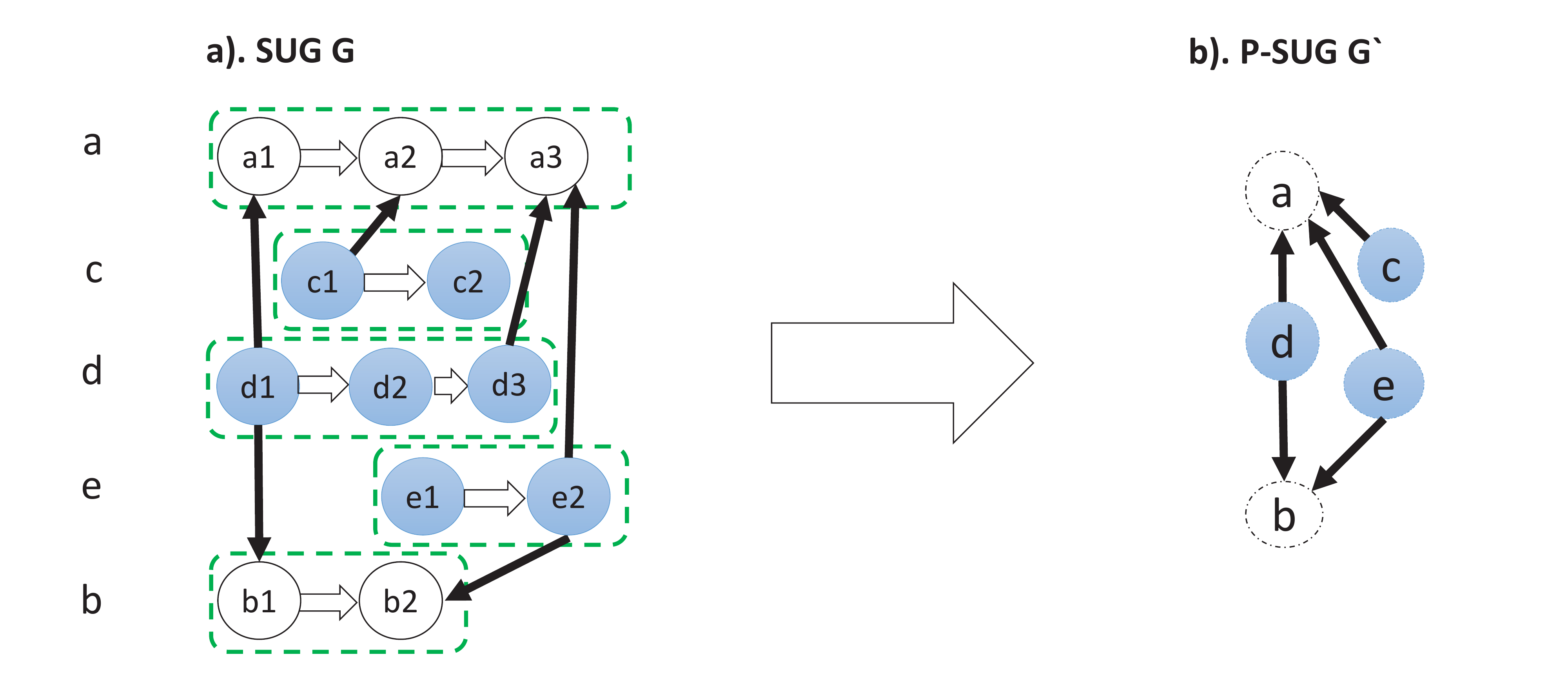}
\caption{Illustrative example of a SUG being reduced to a P-SUG. Let Figure 5(a). shows a typical \SUG~$G$ with respective \Family s annotated. Figure 5(b) is the \pSUG~$G^{\prime}$.} 

\label{fig:famCo}
\end{figure*}

We use the transitive closure properties to define a \Family~set. Hence, dependency evolution can be determined through transitive update-relations such as: $\family(x)\equiv\{y|y\overset{+}\Rightarrow x\ \vee x\overset{+}\Rightarrow y\ \vee x = y \} $ where $a\overset{+}\Rightarrow b$ is the transitive closure on any update-relation $a \Rightarrow b $. The \textit{name} attribute determines \Family~membership. 

The \pSUG~is an aggregation of related \SUG~nodes into a single node.\footnote{Hence, \pSUG~$G^{\prime} =(N^{\prime}, E^{\prime})$ of an \SUG~$G= (N, E=E_{dep} \cup E_{up}$)  where $ E^{\prime} = \{\family(a)\rightarrow \family(b)|  a,b \in N \wedge a \rightarrow b \in E_{dep} \}$ and N$^{\prime} = \{\family(n)|n \in N\} $} Consider the example in Figure \ref{fig:famCo}. Figure  \ref{fig:famCo}(a) shows a typical \SUG~$G$ with respective \Family s annotated. Figure  \ref{fig:famCo}(b) depicts the related P-SUG~$G^{\prime}$. 

To differentiate projects, we use the $\parallel S \parallel$ operator to represent the number of different \Family~in a set of nodes in S. Hence, $ \parallel S \parallel \equiv | \{ \family(s)| s \in S\}|$. For example in Figure \ref{fig:SUG}, suppose $S = \{a1,a2,x1\}$ where $\parallel S\parallel=|\{\family(a1),\family(a2),\family(x1)\}|=|\{\{a1,a2\},\{x1\}\}|= 2$. The \pSUG~variety is used to this extent.

\begin{mydef}
\textit{Variety represents the number of different \Family s that \use~on a project release}
\end{mydef}

\begin{equation}
\variety(u) \equiv \parallel \useBy(u)\parallel\\
\end{equation}

In Figure \ref{fig:SUG} we observe that node $x1$ is used by node related to $\family(a1)$ and $\family(q1)$. Hence, variety is 2. Formally, 
$\variety(x1) = \parallel \{a1,q1,q2\} \parallel  \\ = | \{\family(a1), \family(q1),\family(q2)\} | = 2$. 

\section{\SUG~metrics and visualizations}
Following on from Section \ref{sec:motivate}, in this section we introduce our metrics related to \textit{adoption-diffusion} and  \textit{\couse~pairing} metrics and visualizations.


\subsection{Adoption-diffusion}
As an extension on our work on Library Dependency Plots \cite{2014VISSOFTKula}, we introduce Diffusion Plots (DP). For any project releases, DPs allow us to be able to plot and track both popularity and variety at any given point in time $t$, such that $\pop_{t}(x)$ and $variety_{t}(x)$ for a \SUG~node $x$. For popularity, we plot the number of project releases that depend on a particular release of a \Family. Conversely in the variety plot, we track the number of \Family s that are dependent on a specific release.

The DPs provide a temporal means to evaluate popularity and the adoptive behavior nature. DPs plot both the $\pop_{t}$ and corresponding $variety_{t}$ on a \SUG. We use the plots to understand the adoption and diffusion at both the \Family~and project release levels. Particularly interesting is the temporal \textit{superseding point} (ss point), which is the time at which the point where one project release popularity overtakes another release. DPs also provide a visual analysis of the steepness of the curve; when the curve halts, and when the curve is \textit{superseded} by a successive release curve.

\begin{figure*}
\centering
\includegraphics[width=1.2\textwidth]{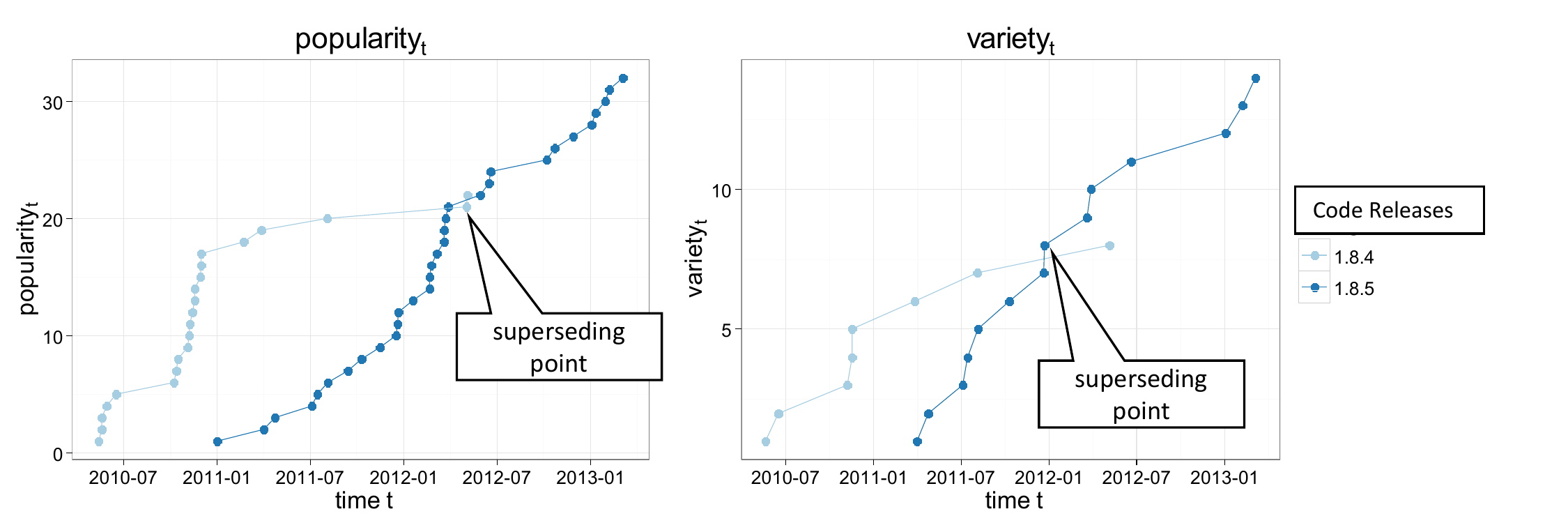}
\caption{A simple example of the adoption-diffusion plot for maven \textsc{mockito-core} \Family~release \texttt{1.8.4} and \texttt{1.8.5}}
\label{fig:DPExample}
\end{figure*}

 In Figure \ref{fig:DPExample} we show an example DP of the \textsc{mockito-core} \Family~from the Maven super repository. For illustration purposes --and to simplify the curve-- this DP only shows two releases. Note the crossing of lines, which is described as the  \textit{superseding point} where $mockito-core_{1.8.5}$ succeeds $mockito-core_{1.8.4}$ in both  $popularity_{t}$ \texttt{(2012-6)} and $variety_{t}$ \texttt{(2011-12)}. In both cases, we conclude that $mockito-core_{1.8.5}$ is the more dominant project release version.

\begin{figure*}
\centering
\includegraphics[width=.8\textwidth]{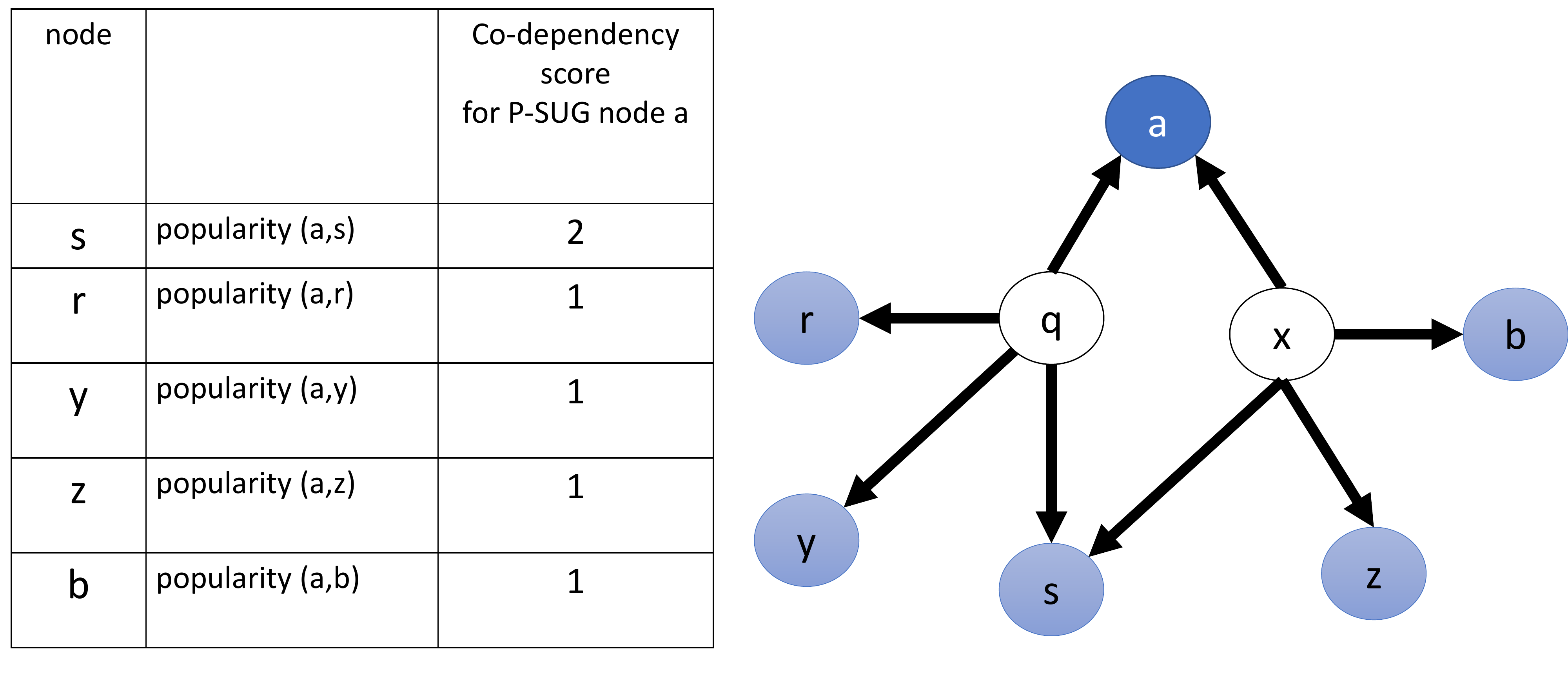}
\caption{Example of the score for the node a. In this example, node $s$ has the highest score in a top 5 listing.}
\label{fig:wild}
\end{figure*}

\subsection{\Couse}
For both \SUG~and P-SUG, the \couse~of two nodes is used to establish a pairing between the nodes. It is defined with an extension of popularity of the note. Popularity of any pair of nodes (denoted by popular($u$, $v$)) is defined by the number of common reverse dependencies (\useBy). Formally, $\pop(u,v)\equiv |\useBy(u) \cap \useBy(v)|$.
We say $u$ and $v$ as \textit{\couse~pairs} if $\pop(u,v)\geq 1 $.
For example, for the P-SUG in Figure \ref{fig:wild}, $a$ and $b$, $a$ and $s$, ..., are examples of \couse~pairs.  
 We propose three types of \couse~pairing: 

\begin{itemize}

\item\textbf{\pSUG~pairs}. We use the popularity of two \pSUG~nodes. Further exploration of the \Family~pairs would lead to release pairs of those respective \Family s.

\end{itemize}

\begin{itemize}
\item\textbf{\SUG~release pairs}. Once interested \Family~pairs are identified, maintainers next decide on popular release combinations. To this end, we use popularity of two nodes on the \SUG~to establish \couse~release pairs.
\end{itemize}

%

\section{Evaluation}
\label{sec:eval}
\subsection{Research Questions}
To evaluate we modeled dependency reuse within real world super repositories. The goal of the evaluation was to answer the following questions:

\begin{itemize}
\item \textbf{RQ1} \RqOne With this question apply our approach to a real world repository to demonstrate practical use cases.
\end{itemize}

 
Secondly, we extend to show how the \SUG~can be used to compare two different super repositories. Thus:

\begin{itemize}
\item \textbf{RQ2} \RqTwo With this research questions we want to demonstrate how the \textit{co-dependency} metric can be leveraged to compare super repositories. 
\end{itemize}

\subsection{Research Method}
For the first research question, the research method is by empirical study of a typical super repository. Then, through use cases, we demonstrate usefulness of each metrics and visualizations. For the second method, using the maven \SUG~generated in the first research questions, we compare the dependencies with a sample of real applications that exist in another super repository. Specifically, we statistically compare dependencies between these different repositories.


\paragraph{Research Method for RQ1} 

For the first research question, we first provide a detailed description of the construction of a \SUG, including the node, edges and attributes definitions and statistics. Next we measure the reuse that occurs within the \SUG. To understand the reuse within a \SUG, we measure how many \Family s are being used internally. Thus, for each \SUG~$U = \{N,E\}$:

\begin{equation}
reuse = \parallel \bigcup\limits_{n \in N} \useBy (n) \parallel
\end{equation}

We use a total of nine popular libraries from the built \SUG~in our case studies, each used in a different scenario to illustrate practicality of our approach. For the \textit{adoption-diffusion} metrics, we use both the \textit{popularity$_t$} and  \textit{variety$_t$} plots. 

In regards to the \textit{co-dependency} metrics, we introduce a heat map style visualization. Using the \pSUG~popularity, we utilize a heat map style with color intensity function to plot popular pair frequency counts.  For the \pSUG~project pairs,  we define \textit{intensity}\footnote{$intensity(\family(x),\family(y),I) = \frac{\pop(\family(x),\family(y))}{\operatorname*{max} \limits_{\substack{i,j\in I\\i \neq j}} \left( \pop(\family(i),\family(j))\right)}$ where the function $max()$ returns the most frequent counts of pairings between \family(x) and \family(y).} as a normalized frequency count of popular pairs with 1 representing the most popular and 0 where no \couse~exists (we use shading to representing the intensity). The \pSUG~Pair Plots serve as a guide for developers to determine the strength of \couse~between  \Family s.

Similarly for the \pSUG~project pairs, the \SUG~Release Pair Plots use the  \textit{popularity intensity on \SUG~nodes} to identify the most popular pairings. Additionally, the release pair plots include the popularity of a specific version $x$ and any other software unit `outside' $\family(y)$. The \textit{outside}\footnote{$outside(x,y)=  \sum \limits_{\substack{n \in U\\\family(n) \neq \family(x)\\\family(n) \neq \family(y)}} popularity(x,n) $} pairs gauge relative popularity of alternative combinations. For example \texttt{outside(x,y)=2} means that there exists 2 \couse~relations with $x$ that are not related to $\family(y)$.  It is plotted at the end of the respective x and y axis of the release pair plots.  From the \Family~pairing (\pSUG), popularity on the \SUG~is used to determine release pairs between two \Family s.


\paragraph{Research Method for RQ2} 
In response to RQ2, we want to quantitatively measure how much common co-dependencies exist between different super repositories. We are interested in comparing the \couse~scores listings generated from the one \SUG~with the actually co-dependencies that exist in another. Suppose there are two super repositories, $SUG_a$ and $SUG_b$. Then,  a system $s$ from $SUG_b$ has a set of library dependencies $\{l_1, l_2, ... , l_n\}$. Thus, for each library $l_i$ that is used,  we compute \SUG~\couse~score from $SUG_a$ and return the top 10 highest scores. We then compute the accuracy of the top 10 list with the ratio of the rest of libraries $l_j$ appearing in this list over the set of libraries in $s$. This method, \textit{top-k accuracy}, is a popular method of evaluation for accuracy \footnote{The function \textit{sysMatch(\textit{\couse}(x))} to determine if at least 1 matches or 0 likewise. Formally: 

\begin{equation}
\textit{accuracy}(P)= \frac{\sum_{r \in P}sysMatch(\textit{\couse}(r) )}{|P|}\times 100 \%
\end{equation}

where P is a set of library dependencies related to one system} \cite{Bacchelli2013}, \cite{pick2015}, \cite{Thung2013}.

\subsection{Dataset}
\label{sec:maven}
For RQ1, we will model the \textsc{Maven} super repository. Maven Central is a specialized library hosting super repository that hosts many JVM project artefacts. Most projects in this super repository are open-source Java, Scala or Clojure libraries (referred to as artefacts). Recently the Maven libraries have been gaining widespread usage do to dependency management tools such as maven and gradle.  We conducted our experiments on a local offline copy of the super repository, which was last updated Feb 2015. 

In our use case we employ nine popular maven libraries. For adoption-diffusion, we use \textsc{Commons-lang} a helper utility library and \textsc{Commons-logging}  a java logger helper library. For the \pSUG~pairs we selected eight popular Maven Apache Commons libraries (logging, lang, dbcp, collections, codec, cli and beanutils) to demonstrate the different \SUG~metrics. Then for the co-dependency release plots, we depict \SUG~release pair plots between \textsc{asm}, \textsc{commons-io}, \textsc{commons-logging} and \textsc{commons-lang} respective libraries.

For RQ2, we use the Maven  \SUG~generated from the RQ1 with a randomly  selected a sample set of 100 systems randomly collected from GitHub\footnote{As our implementation uses R for the statistical analysis, we used the\texttt{ sample()} package to randomly select the systems from a list of over 500 systems} to form the second \SUG. All tools, scripts, data and results are available from the paper's replication package at: \sloppypar{\url{http://sel.ist.osaka-u.ac.jp/people/raula-k/SUG/indexNew.html}}.

\subsection{Construction of the \SUG}
For the Maven super repository, we construct the \SUG~from the POM configuration file. Every project in the Maven repository includes a Project Object Model file (\ie{} \texttt{POM.xml}), that describes the project's configuration meta-data ---including its compile-time dependencies\footnote{Refer to \url{http://maven.apache.org/pom.html} for the data structure }. We customized a tool\footnote{PomWalker: \url{https://github.com/raux/PomWalker}} that implements the maven-model\footnote{maven-model version 3.1.1. Our tool can handle Maven 1.x, 2.x and 3.} parser to extract related \SUG~edges dependency information from all release version of the POM-files in the repository. Similarly encountered by Raemaekers\cite{RaemaekersICSM}, Maven's dependency management mechanism\footnote{\url{http://maven.apache.org/guides/introduction/introduction-to-dependency-mechanism.html}} is rather complex with elements such as transitive and imported POMs. In this study we ignore POM files that reference implicitly (\eg{}  \texttt{<version>\${library.version}</version>}).\footnote{Also references to multiple explicit versions or inconsistent terms such as SNAPSHOT, latest were ignored.}  
Using the formalized model we built the Maven \SUG~$M$ where $M(N,E_{dep}\cup E_{up} )$. Take $x \in N$. We describe each property as follows:
\begin{itemize}
\item \textbf{$E_{dep}$. } The \texttt{<dependency>} attribute of the \texttt{POM.xml} explicitly references the use relation between artefacts. At this stage, we do not resolve transitive dependencies.

\item \textbf{$E_{up}$.} The \texttt{<version>} attribute of the \texttt{POM.xml} explicitly references the release version of an artefact. Using the $time$ attribute of the node, we then determine the order of nodes within a \Family.
\item \textbf{$x.name$.} The \texttt{<artifactId>} was originally used, however it was found in many cases to be too generic. The concatenation of \texttt{<groupId>} produced a more unique \Family~separation. 
\item \textbf{$x.release$.} The \texttt{<version>} attribute of the \texttt{POM.xml} explicitly references the release version of an artefact.
\item \textbf{$x.time$.} The time-stamp of when the artifact (jar file) was uploaded into Maven was used to extract time of the node.
\end{itemize}

A downside of using \texttt{<groupId>} as the name attribute, is that common \Family s are lost if they have moved domain (\ie~changed\texttt{<groupId>}). An example is when the \textsc{findbugs} library change groupID from \texttt{<net.sourceforge.findbugs>} to \texttt{<com.google.code.findbugs>}. Although our tool is unable to resolve explicit references, it is able to handle inheritance attributes of Super POM. Through the \texttt{Dependency Management} attribute the parent and child poms files were resolved.  

\section{Results}
\label{sec:rq1:construct}
\begin{table*}
\begin{center}
\caption{\SUG~Statistics for the Maven Artifacts Packages }
\label{tab: SUGStats}
\begin{tabular}{lcccccc}
\hline
\multirow{1}{*}{\textbf{}}
& \multirow{1}{*}{\textbf{Maven}}\\
\textbf{Time Period} & 2005-11-03 to 2013-11-24 \\
\textbf{\# of nodes}  & 188,951\\
\textbf{\# of \Family s }&  6,374 \\
\textbf{\SUG~node reuse}  & 5,146  \\ \hline
\end{tabular}
\end{center}
\end{table*}

\begin{table}
\begin{center}
\caption{Summary Statistics for \pSUG~popularity}
\label{tab: popStats}
\begin{tabular}{lcccc}
\hline
\multirow{1}{*}{\textbf{}}
& \multirow{1}{*}{\textbf{Maven}}
\\
 Min & 1  \\
 1st Quartile & 2  \\
 Median & 6 \\
 Mean & 38.8 \\
 3rd Quartile & 20\\
 Max & 1016  (\textsc{Junit})\\ \hline
\end{tabular}
\end{center}
\end{table}

Table \ref{tab: SUGStats} details the data mined for the experiment. For our \SUG, our tools were able to mine and generate 188,951 Maven nodes, spanning across 9 years. Independent software units (\ie without use relation edges) were not included in \SUG s. The \SUG s were built from the dates shown in Table \ref{tab: SUGStats}. Table \ref{tab: SUGStats} details the Maven \SUG~statistics. The Maven \SUG~indicates more reuse within super repository (5,146 \Family s used by 6,374 \Family s). This result is typical as most Maven artifacts comprise of libraries or frameworks that may depend on multiple libraries. To determine popularity of a \Family, we apply the popularity function on a \pSUG. Hence, from the Maven, we derive their respective \pSUG s with a \Family~by \Family~dependency-relation. The statistical summary of this \pSUG~popularity distribution for Maven is presented in Table \ref{tab: popStats}. The testing library \textsc{Junit} is the most popular dependency. 

\begin{hassanbox}
\label{insight:RQ1Construction}
Using the dependencies defined in the pom.xml  of each system, we constructed an \SUG~with 188,951 nodes and 6,374 projects, in which 5,146 were reused as libraries.
\end{hassanbox}

\begin{figure*}
\centering
\subfigure[$\textsc{Commons-lang}$]{\label{fig:dbcp}
\includegraphics[width=1.2\textwidth,clip, trim={0cm 0cm 0cm 0cm}]{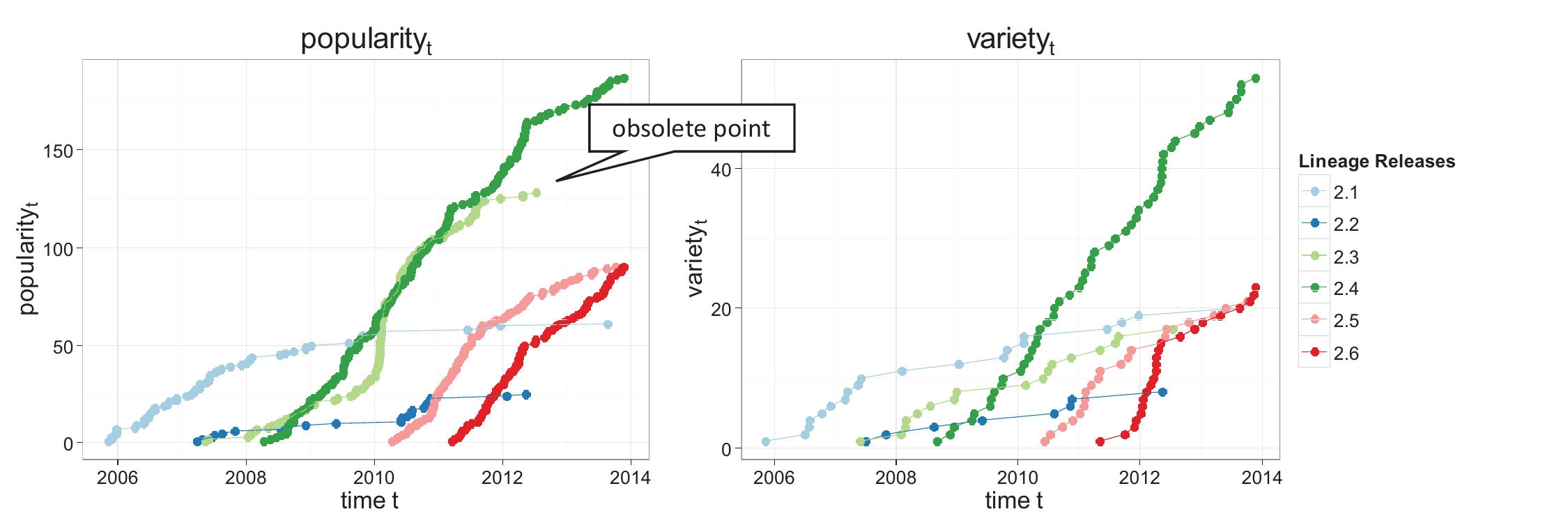}
}
\subfigure[$\textsc{Commons-loggings}$]{\label{fig:loggings}%
\includegraphics[width=1.2\textwidth,clip, trim={0cm 0cm 0cm 0cm}]{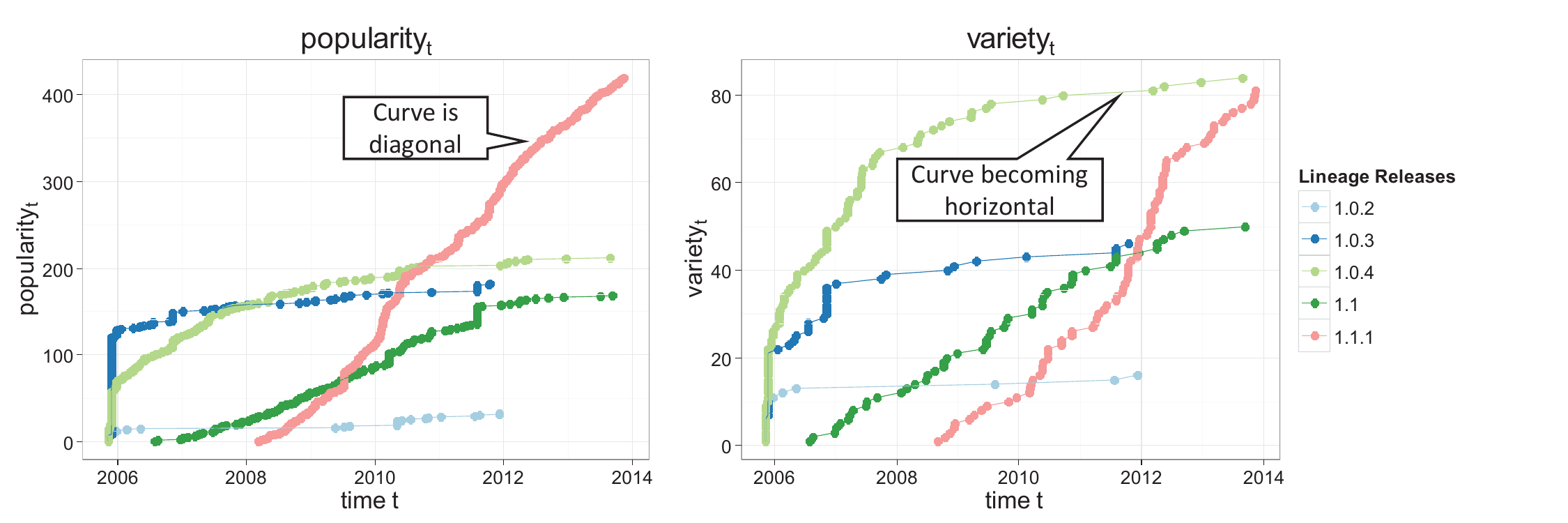}
}
\caption{Diffusion Plots for selected Maven \Family s. The left hand depicts the $popularity_t$ while the right shows the $variety_t$ for their respective releases.}
\label{fig:DDPMaven}
\end{figure*}

%

\subsection*{Adoption-diffusion (Best library version at any point in time?)}

As seen in Figure \ref{fig:dbcp}, \textsc{Commons-lang$_{2.4}$} (dark green)  although older is the most popular release.  Since the last superseding point of \textsc{Commons-lang$_{2.4}$} and \textsc{Commons-lang$_{2.3}$} is between 2010 and 2011, none of the latter versions have been able to supersede it. This is consistent in both $\pop_t$ and $variety_t$ plots. 

The steepness of the curve can indicate strength of popularity. For instance in the $variety_t$ plot of Figure \ref{fig:loggings}, we observe that \textsc{Commons-logging$_{1.0.4}$} has the most variety at any point in time. However, closely looking at its curve (light green), popularity has probably peaked with the curve almost horizontal. \textsc{Commons-logging$_{1.1.1}$} (pink), however, adopts a more diagonal curve, hinting future adoptions could follow this trend. Note that the predecessor \textsc{Commons-lang$_{2.4}$} (dark green) in Figure \ref{fig:dbcp} is still adopted beyond the \textsc{Commons-lang$_{2.3}$} obsolete point, making it more successful.

 As shown in  Figure \ref{fig:loggings}, significant differences between $popularity_t$ and the corresponding $variety_t$ indicate \Family s with abnormally high releases depend on this specific \Family~release. $popularity_t$ depicts \textsc{Commons-logging$_{1.1.1}$} clearly as the popular version, however the corresponding $variety_t$ suggest \textsc{Commons-logging$_{1.0.4}$} is still as dominant across systems. 

\begin{hassanbox}
\label{insight:RQ1Diffusion1}
Older versions releases may still be heavily used in the super repository. The steepness of the curve hints of potential popularity. Popular but saturating (horizontal) plots may indicate the version becoming obsolete. We can compare between the popularity$_t$ and variety$_t$ to distinguish between popularity within one system verses across systems.
\end{hassanbox}

\begin{figure*}
\centering
\includegraphics[width=.5\textwidth]{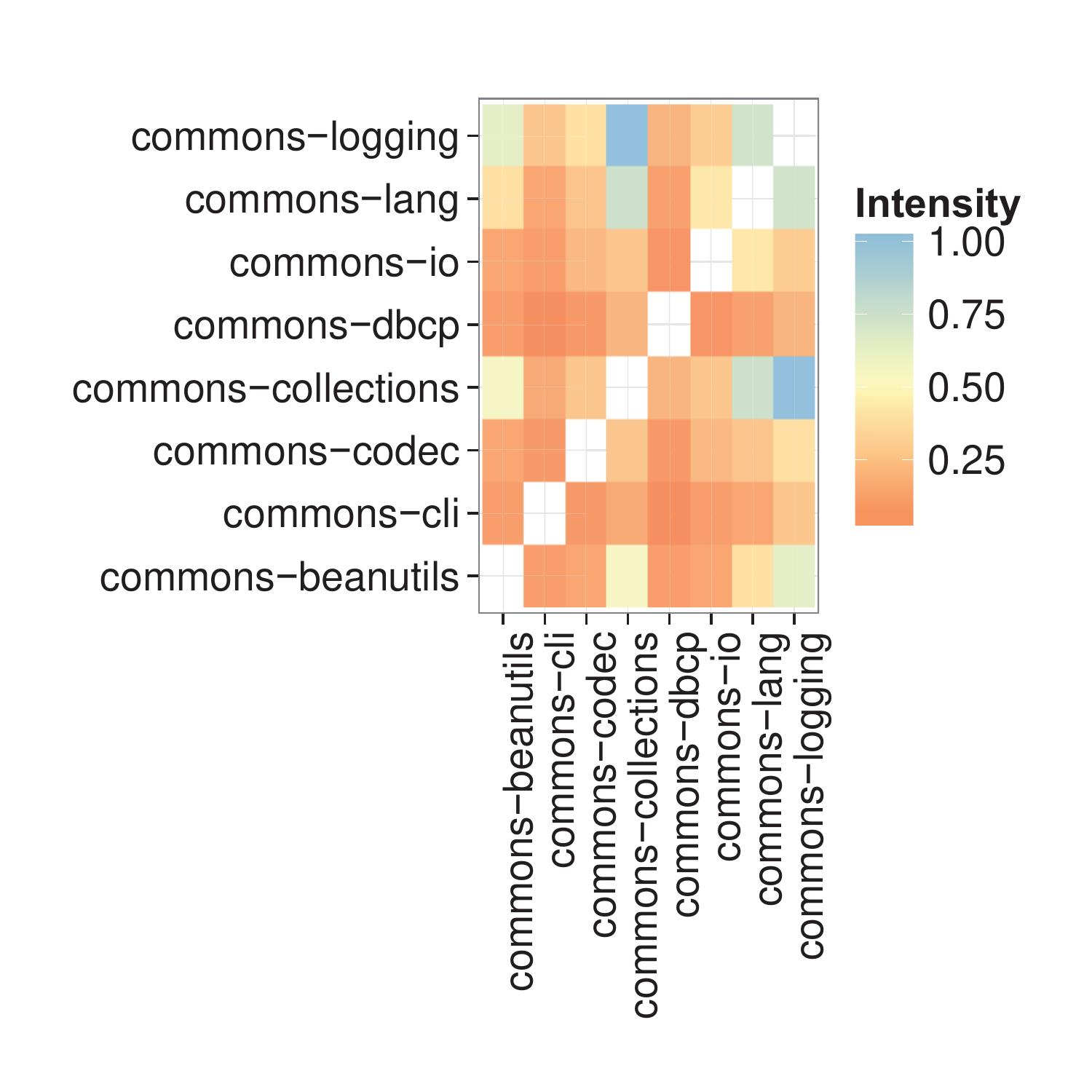}
\caption{P-SUG~pairing plot of 8 Apache Commons artifacts}
\label{fig:familyPairs}
\end{figure*}

\subsection*{\Couse~Pairing (Best pairing of library versions)} 
\label{sec:releasePair}

Figure \ref{fig:familyPairs} depicts the pairing of eight selected Maven Apache Commons artifacts built for java. From the matrix, it is observed that the most popular pairing is between \textsc{commons-logging} and \textsc{commons-collections}. Thus, the recommendation is that for a system using one of the library, it is worth considering the other for adoption.

\begin{figure*}
\centering
\subfigure[\textsc{Asm} vs. \textsc{commons-Logging}.]{\label{fig:CollectLang}
\includegraphics[width=.45\textwidth,clip, trim={0cm 0cm 0cm 0cm}]{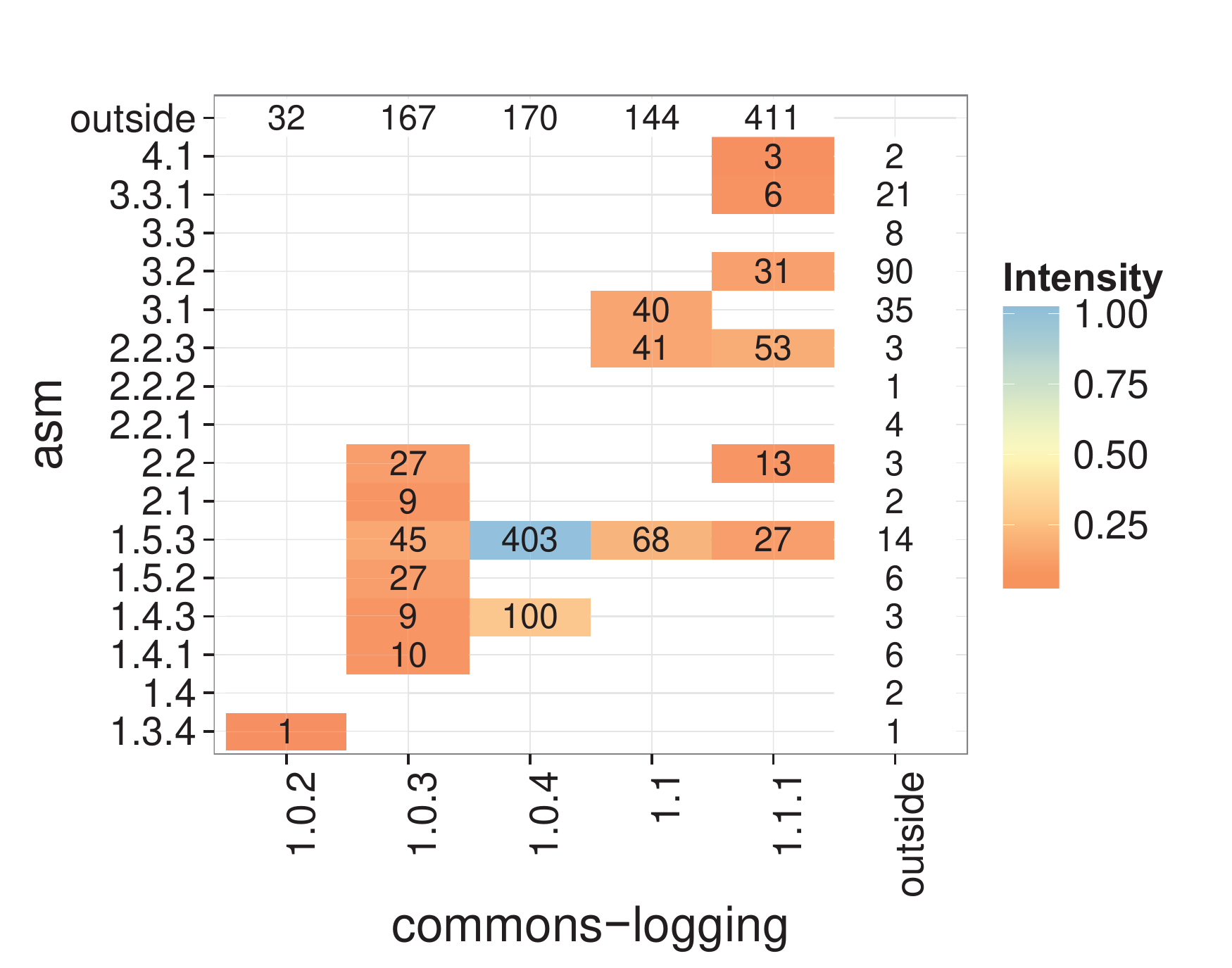}
}
\subfigure[\textsc{Asm} vs. \textsc{Commons-io}.]{
\label{fig:CollectIo}%
\includegraphics[width=.45\textwidth,clip, trim={0cm 0cm 0cm 0cm}]{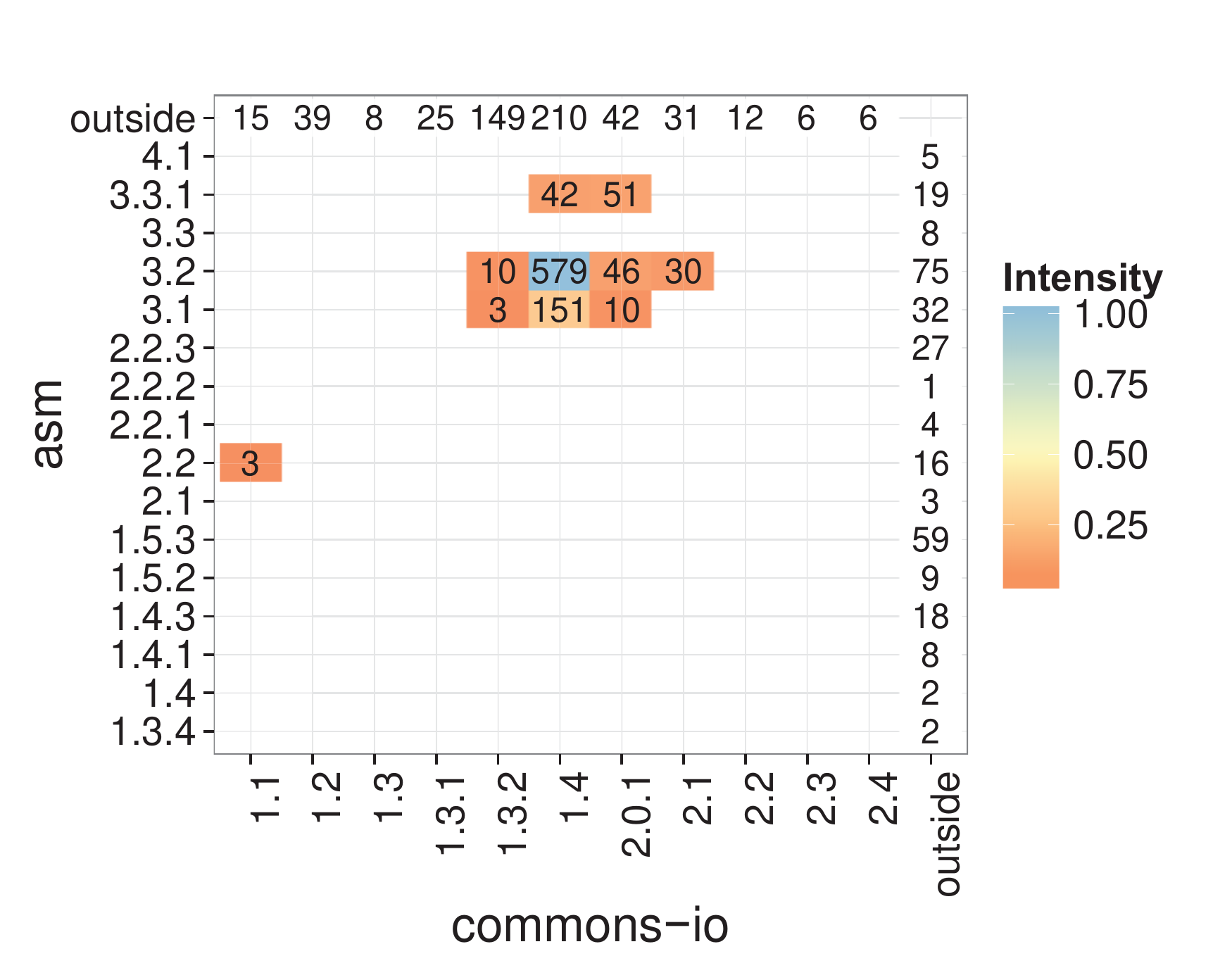}
}
\subfigure[\textsc{Commons-Lang} vs. \textsc{Commons-Logging}.]{\label{fig:LangLog}%
\includegraphics[width=.45\textwidth,clip, trim={0cm 0cm 0cm 0cm}]{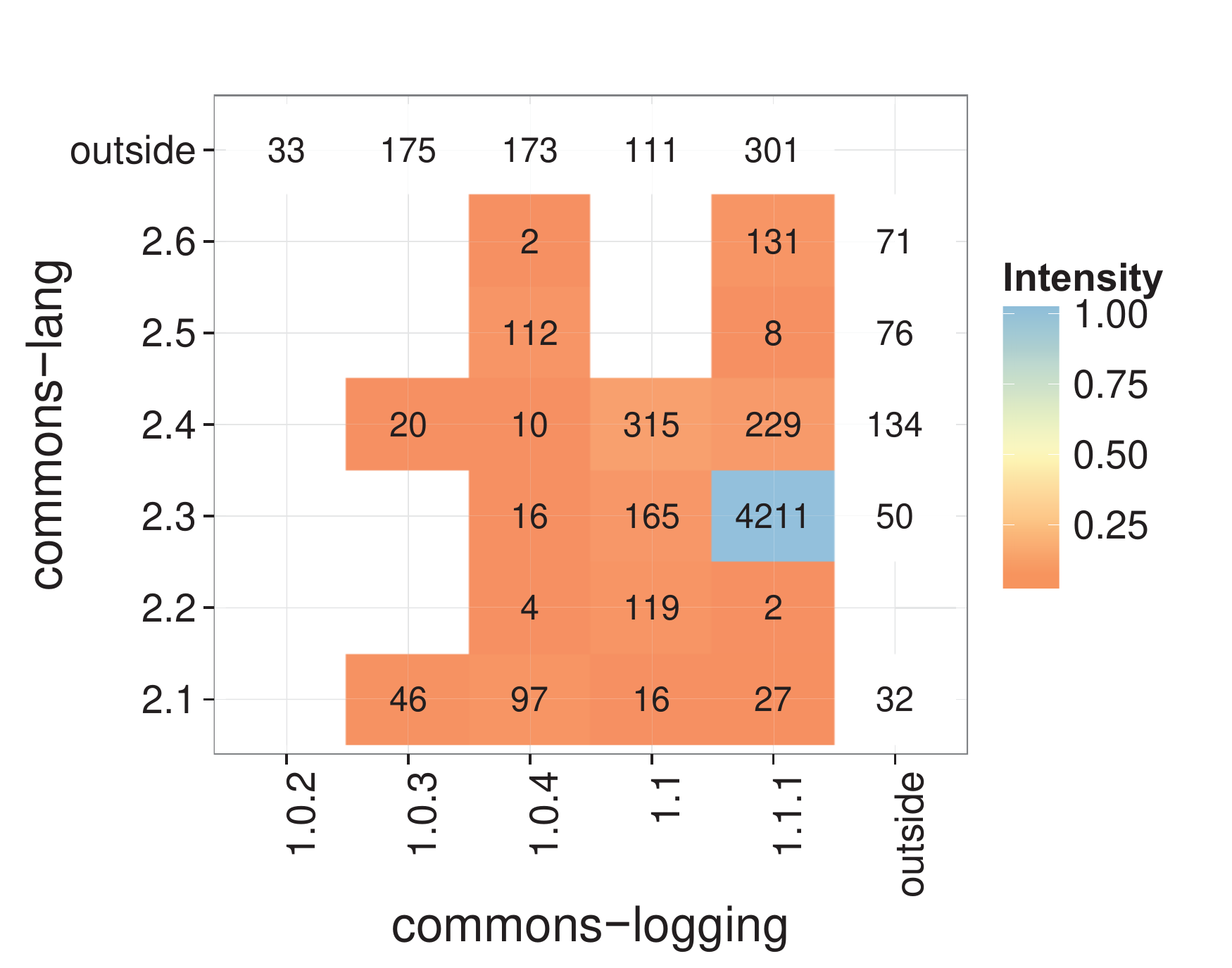}
}
\subfigure[\textsc{Commons-Lang} vs. \textsc{Commons-Io}.]{\label{fig:LangIo}%
\includegraphics[width=.45\textwidth,clip, trim={0cm 0cm 0cm 0cm}]{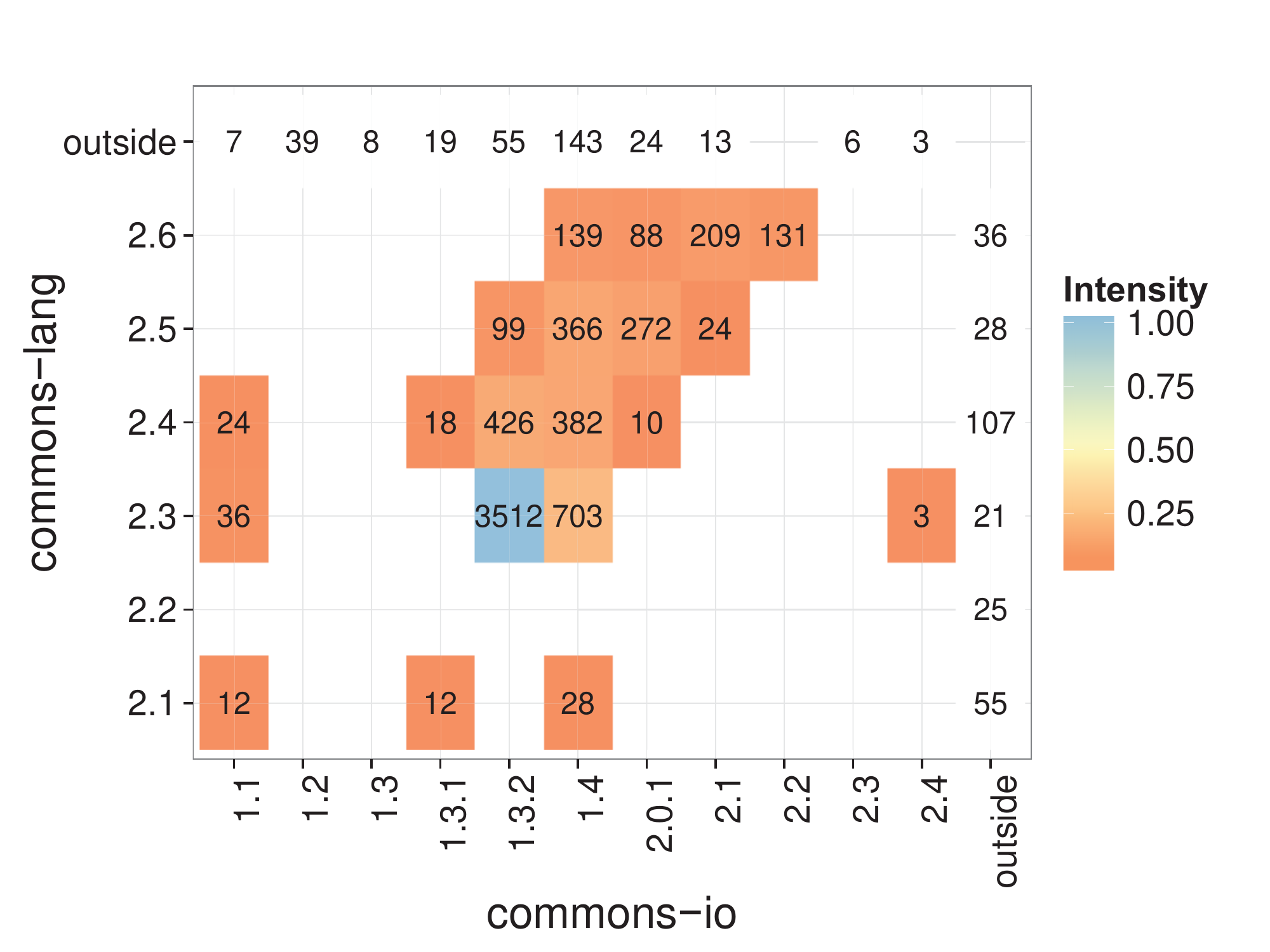}
}
\caption{Example of \SUG~release~plots for Maven Artifacts (a,b,c,d). }
\label{fig:CoExistMaven}
\end{figure*}



\begin{hassanbox}
\label{insight:RQ1Codependents}
The co-dependency P-SUG pairing plot recommends that for a system using one of the library, it is worth considering this other library for adoption.
\end{hassanbox}

Figure \ref{fig:CollectLang}, \ref{fig:CollectIo}, \ref{fig:LangLog} and \ref{fig:LangIo}  depicts the SUG release pair plots between Maven's \textsc{asm}, \textsc{commons-io}, \textsc{commons-logging}   and \textsc{commons-lang}. Different to the \pSUG~\Family~pairs, the popularity is annotated at each pairing point. In Figure \ref{fig:CollectIo}, the popularity of pairing Maven \textsc{commons-io}$_{1.4}$ and \textsc{asm}$_{3.2}$  (popularity of 579) is greater than both `outside' \textsc{commons-io}$_{1.4}$ (popularity of 210) and \textsc{asm}$_{3.2}$ (popularity of 75), validating it as a very popular \couse~relation.

Latent migration patterns such as the use of older versions are clearly apparent in the \SUG~release pair plots. For instance Figures \ref{fig:LangLog} and \ref{fig:LangIo} depict \textsc{Commons-logging} (particularly the newer \textsc{Commons-logging}$_{1.1.1}$) has \couse~with older versions of \textsc{Commons-lang} (such as \textsc{Commons-lang}$_{2.1}$), whereas newer versions of \textsc{Commons-io}(versions 2.0.1 onwards) tend have \couse~with newer versions of the \textsc{commons-lang} (versions 2.5 onwards)  library.

\begin{hassanbox}
\label{insight:RQ1Codependents}
The SUG release plots depict co-evolution patterns between versions of two libraries. For any two libraries, we can deduce 1.) popular combinations, 2.) compare to outside combinations and  3.) latent migration patterns.
\end{hassanbox}

\begin{figure*}
\centering
\includegraphics[width=0.6\textwidth,clip, trim={0cm 0cm 0cm 0cm}]
{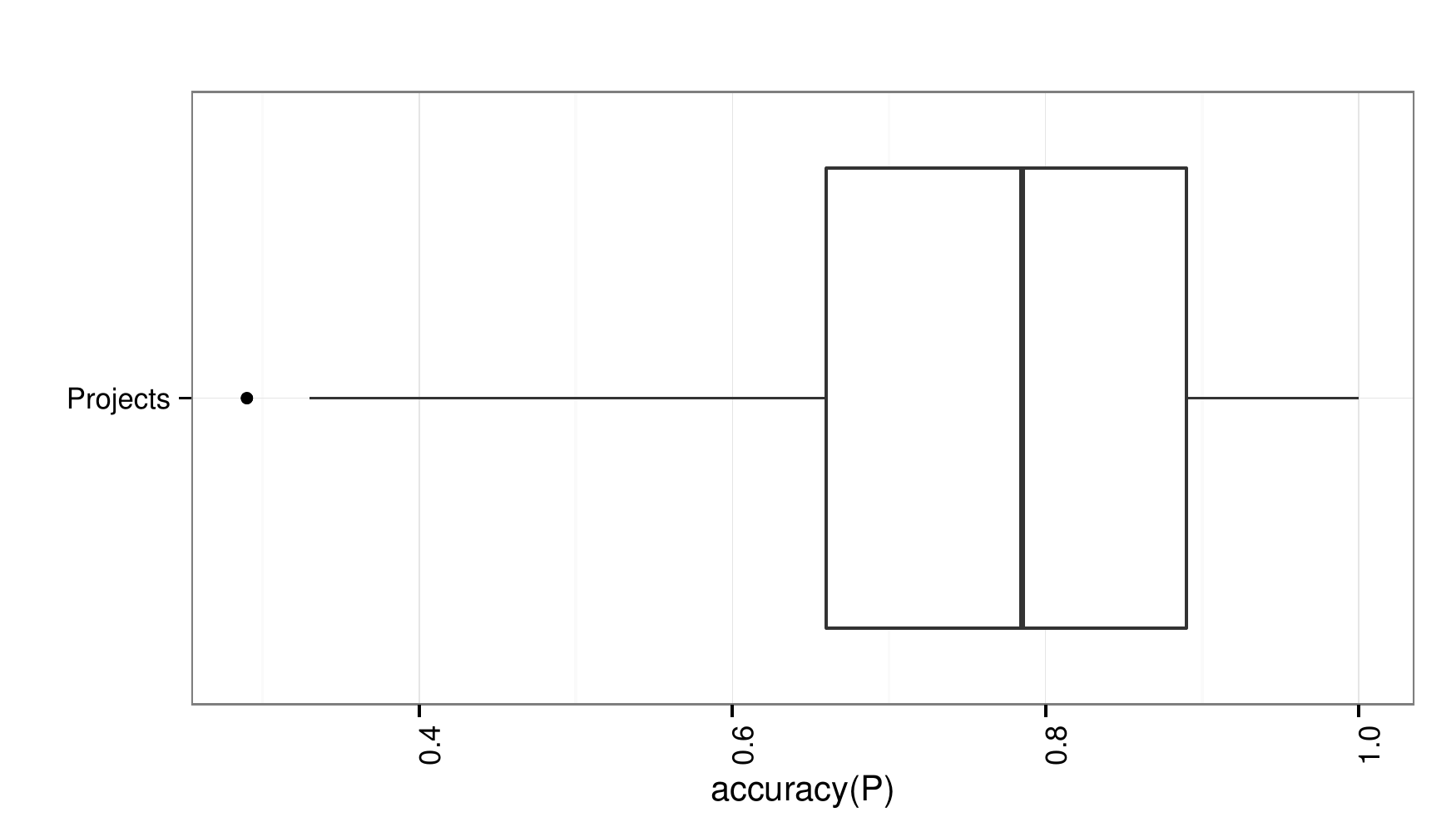}
\caption{We show the accuracy(P) (n\%) of common dependencies between maven and github systems. The result suggests that many of the projects in github could be from the maven super repository or adopt similar dependency habits.}
\label{fig:RQ2github}
\end{figure*}


\subsection{Dependencies across Super Repositories}


Depicted in Figure \ref{fig:RQ2github}, the results of the experiment show a median of 79\% of the \couse~scores on the Maven \SUG~contained actual libraries in GitHub. We think that 79\% is relatively high, implying \couse~relations in GitHub are very similar to those implemented in Maven. A obvious conclusion could be that many Maven libraries exist in Github, thus making them very similar super repositories. Full details of all 100 random system statistics are available at \sloppypar{ \url{http://sel.ist.osaka-u.ac.jp/people/raula-k/SUG/rq2/accuracyR.csv}}. 


\begin{hassanbox}
\label{insight:RQ2Codependents}
We leverage the \SUG~co-dependency to show that Maven and GitHub systems share similar dependencies, indicating overlap between the super repositories. 
\end{hassanbox}

\section{Discussion}

\subsection{Study Implications}
Advancements in data storage and mining repository techniques and tools, make possible the study of popularity and `follow the crowd' approaches. Other related work have based their work on trend and pure popularity of usage, the results show that our recommendation are an extension of the simple popularity. We believe the examples demonstrate the merits of using the SUG to recommend: 1.) the diffusion plot could be used to predict future popularity or obsolete of library versions. 2.) From a set of candidate libraries, the P-SUG can recommend the best combination pairing of libraries. 3.) the SUG release plots not only shows the popular combinations between two libraries, but also what version is popular outside the combination. Additionally the co-dependency evolution patterns depict if the two libraries have been evolving at the same time. 

The \SUG~can be used to compare dependencies between two system. In our empirical study, we compare a producer type super repository (github) against its library super repository (maven). Our results suggest overlapping, thus empirically that most github projects are either maven libraries or follow similar dependency habits as their counterparts in maven. The result is significant as it shows that their is common knowledge among both maintainers of applications and libraries.

\subsection{\SUG~Extensions}
Our \SUG~model is designed to rely on the dependency chains but differs from the Graph cyclic based approaches such as ranking (such as page ranking), reference counting and component ranking is common for measuring popularity on graph based models \cite{Teyton2012}, \cite{ReuseInoue}. The current graph modeled structure allows for faster and scalable querying, which we utilize for the diffusion and \couse~metrics.  


We envision the \SUG~as a foundation in which many other features can be built. As we study more systems, we will consider integrating `containment' and `transitive' concepts of object-oriented software into the \SUG. We also plan to address issues of authentication of the \texttt{name} attribute. We plan to expand beyond the name attribute for \Family~classifications, by incorporating more sophisticated techniques and tools used in `code clone' such as code clone detection \cite{2007Roy}, \cite{KamiyaCC} and  `origin' analysis  \cite{GodfreyTSE}, \cite{DaviesMSR2011} to determine a common \Family. Another complex but useful operation that was not presented in this paper is the tracing of systems that have abandoned or dropped a library dependency.




\subsection{Threats to Validity}
\subsubsection*{Internal}
The main threat to validity is the real-world evaluation by maintainers. We have been working closely with system integration industrial partners to develop and test our visualizations. We argue though our examples are sufficient to demonstrate possible library recommendations. In this study, we used the pom.xml attributes to build the \SUG. The abstract nature of the \SUG~allows for incorporation of other programming languages which provide their own library hosting repository. Therefore, we believe the \SUG~to be a universal approach for any type of super repository. A threat to RQ2 would be the rank list size and using other measures such as the Mean Reciprocal Rank. We consider our sample size is 100 real systems and a rank list of 10 libraries to be sufficient for the purpose of the study. 


\subsubsection*{External}
Our datasets only includes information about dependencies that are explicitly stated in project configuration files, such as the Maven \texttt{POM} configuration files. It does not take into account reuse such as copy-and-paste and clone-and-own. Although gauging dependencies by the configuration file only provides for a sample of the actual reuse, we believe this is sufficient to give an impression of  trends within each universe. We understand that our data and analysis are dependent on the tools and analysis techniques. Threats include parsing techniques. However, we believe that our samples are large enough to be representative of the real world.

\section{Related Work}
\subsection{Popularity Metrics}
Studying library usage in terms of absolute popularity is not a new concept.
Holmes et al.\ appeal to popularity as the main indicator to identify libraries of interest~\cite{HolmesW07}. Eisenberg et al.\ improve navigation through a library's  structure using the popularity of its elements to scale their depiction~\cite{EisenbergEtAl10a}.
De Roover et al.\ explore library popularity in terms of source-level usage patterns~\cite{2013:icpc:deroover}.
Popularity over time has received less attention.
Mileva et al.\ study popularity over time to identify the most commonly used library versions~\cite{MilevaEtAl09}. Follow-up work applies the theory of diffusion to identify and predict version usage trends~\cite{MilevaDZ10}. Similar to our diffusion work, Bloemen et al.~\cite{Bloemen:2014} explored the diffusion of Gentoo packages. Using the Bass Diffusion Model, they modeled the diffusion of Gentoo packages over time. Other related work includes the `library migration graphs' of Teyton et al.\cite{Teyton2012}. Recently Hora introduced apiwave in visualizations to show popularity trends at the API level. \cite{hora2015apiwave}.

Our work extends on popularity for more indepth analysis of the `wisdom of the crowd'. Our study investigates \couse~and diffusion instead of migration. Consequently, our graph implements an incremental approach as opposed to the cyclic migration graph model. 


\subsection{The Software Repository Universe as Ecosystems}

Recently, there has been an increase in research that perceives software systems as  ecosystems. Work such as Bosch \cite{2009Bosch:SPL2009} have studied the transition from Product Lines to an Software Ecosystem approach. German et al.~\cite{2013CSMRGerman} studied the GNU R project as an ecosystem over time. Since the projects inception, the studied found that user-contributed systems have been growing faster than core-systems and identified differences of how they attracted active communities.
Mens et al~\cite{Mens:ECOS}  perform ecological studies of open source software ecosystems with similar results.

Haenni et al.~\cite{2013:wea:haenni} performed a survey to identify the information that developers lack to make decisions about the selection, adoption and co-evolution of upstream and downstream projects in a software ecosystem.

\subsection{Code Search and  Library Recommendation Systems}
 Code search is prominent among research on software reuse with many benefits for system maintainers \cite{2011:Bajracharya}. Examples of available code search engines are google code\footnote{\url{https://code.google.com/}} and black duck open hub \footnote{\url{https://code.openhub.net/}}. Tools such as  Ichi-tracker \cite{Inoue:2012}, Spars \cite{ReuseInoue}, MUDAblue \cite{KawaguchiAPSEC2004-journal2006} and ParserWeb \cite{Thummalapenta:2007:ParserWeb} just a few of the many available search tools that crawl software repositories mining different software attributes and patterns with different intentions. For instance, SpotWeb searches for different library usage patterns while MUDAblue automatically categorizes related software systems. 
 We crawl the super repositories, using mined data to construct our abstract \SUG~models. Differently, our work involves purely popularity metrics to locate through model operations and visualization different \couse~and adoption-diffusion behavior. 
 
Most existing library recommendation work are based on commonly used together patterns at the method level, i.e., API usage patterns at the method level of granularity. Other related work only recommend support for existing libraries in systems, using code examples or linkage to online learning resources. The most related work of recommendation at the library level of granularity is by Thung et al. \cite{Thung2013}. Through Mining Software Repositories (MSR), they  use association rule mining on historic software artifacts to determine commonly used libraries. Inspired by these existing work, we believe the \SUG~model can be leveraged by to expand the current work and provide a means towards better library recommendation systems.

In regard to the \SUG~attributes and properties, there exists many related definitions of software variability and dependency relationships. In Software Product Line Engineering (SPL), terms such as `product' variability has been used extensively \cite{2009Bosch:SPL2009}, \cite{2013SeidlVamos}, \cite{Nonaka09SIGSEe}. In the code clones field, Kim et al. \cite{KimFSE2005} coined clone `genealogies' to track variability between software of similar origins. 
In addition, systems and libraries are not explicitly distinguished.  The \couse~operations on the \SUG~demonstrate more `basic' aspects of the model, although domain specific filtering may be required. Another complex but useful operation that was not presented in this paper is the tracing of systems that have abandoned or dropped a library dependency.

\section{Conclusion and Future Work}
OSS libraries are now prominent in modern software development. With the advent of \textsc{Maven}, \textsc{Sourceforge}, and \textsc{GitHub}, several opportunities have arisen to uncover insights valuable to the management of library dependencies through intelligent super repository mining. In this paper, we presented the \SUG~model as a means to represent, query and visualize different super repositories in a generic manner. Immediate future work focuses on evaluating the \SUG~with actual system maintainers. We are also developing \SUG s prototypes of different super repositories to gain feedback and explore other potential uses of the model. 

Our work is towards empowering maintainers to make more informed decisions about whether or not to update the library dependencies of a system. Combining its ``wisdom-of-the-crowd'' insights with complementary work on compatibility checking of API changes, should give rise to a comprehensive recommendation system for dependency management.
%

\section*{References}





 \bibliographystyle{elsarticle-num} 
 \bibliography{sigproc}

\begin{thebibliography}{10}
\expandafter\ifx\csname url\endcsname\relax
  \def\url#1{\texttt{#1}}\fi
\expandafter\ifx\csname urlprefix\endcsname\relax\def\urlprefix{URL }\fi
\expandafter\ifx\csname href\endcsname\relax
  \def\href#1#2{#2} \def\path#1{#1}\fi

\bibitem{EbertOSS}
C.~Ebert, Open source software in industry, in: IEEE Software, 2008, pp.
  52--53.

\bibitem{HainemannICSR2011}
L.~Hainemann, F.~Deissenboeck, M.~Gleirscher, B.~Hummel, M.~Irlbeck, On the
  extent and nature of software reuse in open source java projects, in:
  Proceedings of the 12th International Conference on Top Productivity Through
  Software Reuse, 2011, pp. 207--222.

\bibitem{springURL}
Spring io, accessed 2015-08-01, \url{https://spring.io/}.

\bibitem{commonsURL}
Apache commons, accessed 2015-08-01, \url{http://commons.apache.org/}.

\bibitem{MavenCentralURL}
The maven central super repository, accessed 2015-08-01,
  \url{http://search.maven.org/}.

\bibitem{sourceforgeURL}
Sourceforge super repository, accessed 2015-08-01,
  \url{http://sourceforge.net/}.

\bibitem{githubURL}
Github super repository, accessed 2015-08-01, \url{https://github.com/}.

\bibitem{Grinter}
R.~E. Grinter, Understanding dependencies: A study of the coordination
  challenges in software development, Ph.D. Thesis. University of California.
  Department of Information and Computer Science.

\bibitem{mattsson1999framework}
M.~Mattsson, J.~Bosch, M.~E. Fayad, Framework integration problems, causes,
  solutions, Communications of the ACM 42~(10) (1999) 80--87.

\bibitem{Fayad:1999}
M.~E. Fayad, D.~C. Schmidt, R.~E. Johnson, Building Application Frameworks:
  Object-oriented Foundations of Framework Design, John Wiley \& Sons, Inc.,
  New York, NY, USA, 1999.

\bibitem{Kula2015}
R.~G. Kula, D.~M. German, T.~Ishio, K.~Inoue, Trusting a library: A study of
  the latency to adopt the latest maven release, in: 22nd IEEE International
  Conference on Software Analysis, Evolution, and Reengineering, SANER 2015,
  Montreal, Canada, March 2-6, 2015, 2015.

\bibitem{Dagenais:2009}
B.~Dagenais, M.~P. Robillard, Semdiff: Analysis and recommendation support for
  api evolution, in: Proceedings of the 31st International Conference on
  Software Engineering, ICSE '09, IEEE Computer Society, Washington, DC, USA,
  2009, pp. 599--602.

\bibitem{clirrrURL}
Clirr tool and library, accessed 2015-08-01,
  \url{http://clirr.sourceforge.net/}.

\bibitem{2014VISSOFTKula}
R.~G. Kula, C.~D. Roover, D.~M. German, T.~Ishio, K.~Inoue, Visualizing the
  evolution of systems and their library dependencies, Proc. of IEEE Work.
  Conf. on Soft. Viz. (VISSOFT).

\bibitem{YanoICPC2015}
Y.~Yano, R.~G. Kula, T.~Ishio, K.~Inoue, Verxcombo: An interactive data
  visualization of popular library version combinations, in: 23rd IEEE
  International Conference on Program Comprehension, ICPC 2015, Firenze, Italy,
  May 18-19, 2015, 2015.

\bibitem{DoI}
E.~M. Rogers, Diffusion of innovations, 5th Edition, Free Press, NY, 2003.

\bibitem{Venkatesh2004}
S.~Chuan-Fong, V.~Alladi, Beyond adoption: Development and application of a
  use-diffusion model, Journal of Marketing.

\bibitem{SulaymanRoR07}
S.~K. Sowe, L.~Angelis, I.~Stamelos, Y.~Manolopoulos, Using repository of
  repositories (rors) to study the growth of f/oss projects: A meta-analysis
  research approach, in: IFIP International Federation for Information
  Processing, 2007.

\bibitem{LunguWCRE07}
M.~Lungu, M.~Lanza, T.~Gîrba, R.~Heeck, Reverse engineering
  super-repositories, in: Work. Conf, Rev. Eng. WCRE07, 2007.

\bibitem{2013CSMRGerman}
D.~M. German, B.~Adams, A.~E. Hassan, The evolution of the r software
  ecosystem, Proc. of European Conf. on Soft. Main. and Reeng. (CSMR2013)
  (2013) 243--252.

\bibitem{2013:wea:haenni}
N.~Haenni, M.~Lungu, N.~Schwarz, O.~Nierstrasz, Categorizing developer
  information needs in software ecosystems, in: Proc. of Int. Work. on Soft.
  Eco. Arch. (WEA13), 2013, pp. 1--5.

\bibitem{Bacchelli2013}
A.~Bacchelli, C.~Bird,
  \href{http://research.microsoft.com/apps/pubs/default.aspx?id=180283}{Expectations,
  outcomes, and challenges of modern code review}, in: Proceedings of the
  International Conference on Software Engineering, IEEE, 2013.
\newline\urlprefix\url{http://research.microsoft.com/apps/pubs/default.aspx?id=180283}

\bibitem{pick2015}
P.~Thongtanunam, C.~Tantithamthavorn, R.~G. Kula, N.~Yoshida, H.~Iida, K.~ichi
  Matsumoto, Who should review my code? a file location-based code-reviewer
  recommendation approach for modern code review, in: 22nd IEEE International
  Conference on Software Analysis, Evolution, and Reengineering, SANER 2015,
  Montreal, Canada, March 2-6, 2015, 2015.

\bibitem{Thung2013}
F.~Thung, D.~Lo, J.~Lawall, Automated library recommendation, in: Reverse
  Engineering (WCRE), 2013 20th Working Conference on, 2013, pp. 182--191.
\newblock \href {http://dx.doi.org/10.1109/WCRE.2013.6671293}
  {\path{doi:10.1109/WCRE.2013.6671293}}.

\bibitem{RaemaekersICSM}
S.~Raemaekers, A.~van Deursen, J.~Visser, Measuring software library stability
  through historical version analysis, in: Proc. of Intl. Comf. Soft. Main.
  (ICSM), 2012, pp. 378--387.

\bibitem{Teyton2012}
C.~Teyton, J.-R. Falleri, X.~Blanc, Mining library migration graphs, in: Proc.
  of. Work. Conf. on Rev. Eng. WCRE2012, 2012, pp. 289--298.

\bibitem{ReuseInoue}
K.~Inoue, R.~Yokomori, T.~Yamamoto, M.~Matsushita, S.~Kusumoto, Ranking
  significance of software components based on use relations, Software
  Engineering, IEEE Trans. 31 (2005) 213--225.

\bibitem{2007Roy}
C.~K. Roy, J.~R. Cordy, A survey on software clone detection research, in:
  Technical Report No. 2007-541,Queen’s University, Canada, 2007.

\bibitem{KamiyaCC}
T.~Kamiya, S.~Kusumoto, K.~Inoue, {CCFinder:} a multilinguistic token-based
  code clone detection system for large scale source code, IEEE Transactions on
  Software Engineering 28~(7) (2002) 654--670.
\newblock \href {http://dx.doi.org/10.1109/TSE.2002.1019480}
  {\path{doi:10.1109/TSE.2002.1019480}}.

\bibitem{GodfreyTSE}
M.~Godfrey, L.~Zou, Using origin analysis to detect merging and splitting of
  source code entities, IEEE Transactions on Software Engineering 31~(2) (2005)
  166--181.

\bibitem{DaviesMSR2011}
J.~Davies, D.~M. German, M.~W. Godfrey, A.~Hindle, Software bertillonage:
  Finding the provenance of an entity, in: Proceedings of the 8th Working
  Conference on Mining Software Repositories, 2011, pp. 183--192.

\bibitem{HolmesW07}
R.~Holmes, R.~J. Walker, Informing {E}clipse {API} production and consumption,
  in: OOPSLA2007, 2007, pp. 70--74.

\bibitem{EisenbergEtAl10a}
D.~S. Eisenberg, J.~Stylos, A.~Faulring, B.~A. Myers, Using association metrics
  to help users navigate {API} documentation, in: VL/HCC2010, 2010, pp. 23--30.

\bibitem{2013:icpc:deroover}
C.~{De Roover}, R.~L\"ammel, E.~Pek, Multi-dimensional exploration of api
  usage, in: Proc. of IEEE Intl. Conf. on Prog. Comp.(ICPC13), 2013.

\bibitem{MilevaEtAl09}
Y.~M. Mileva, V.~Dallmeier, M.~Burger, A.~Zeller, Mining trends of library
  usage, in: ERCIM Workshops, 2009, pp. 57--62.

\bibitem{MilevaDZ10}
Y.~M. Mileva, V.~Dallmeier, A.~Zeller, Mining {API} popularity, in: TAIC PART,
  2010, pp. 173--180.

\bibitem{Bloemen:2014}
R.~Bloemen, C.~Amrit, S.~Kuhlmann, G.~Ord\'{o}\~{n}ez Matamoros, Innovation
  diffusion in open source software: Preliminary analysis of dependency changes
  in the gentoo portage package database, in: Proc. of Work. Conf. on Mining
  Soft. Repo. (MSR2014), 2014, pp. 316--319.

\bibitem{hora2015apiwave}
A.~Hora, M.~T. Valente, apiwave: Keeping track of api popularity and migration,
  in: International Conference on Software Maintenance and Evolution, 2015.

\bibitem{2009Bosch:SPL2009}
J.~Bosch, From software product lines to software ecosystems, in: Proc.of the
  Int Soft. Prod. Line (SPLC '09), 2009, pp. 111--119.

\bibitem{Mens:ECOS}
T.~Mens, M.~Claes, P.~Grosjean, Ecos: Ecological studies of open source
  software ecosystems, in: Soft. Main. Reeng. and Rev. Eng. (CSMR-WCRE), 2014,
  pp. 403--406.

\bibitem{2011:Bajracharya}
S.~Bajracharya, A.~Kuhn, Y.~Ye, Proc. of work. on search-driven dev.: Users,
  infrastructure, tools, and evaluation (suite 2011), in: Proceedings of the
  33rd International Conference on Software Engineering, 2011.

\bibitem{Inoue:2012}
K.~Inoue, Y.~Sasaki, P.~Xia, Y.~Manabe, Where does this code come from and
  where does it go? - integrated code history tracker for open source systems
  -, in: Proc. of Inl Conf. on Soft. Eng., ICSE 2012, IEEE Press, Piscataway,
  NJ, USA, 2012, pp. 331--341.

\bibitem{KawaguchiAPSEC2004-journal2006}
S.~Kawaguchi, P.~K. Garg, M.~Matsushita, K.~Inoue, {MUDABlue:} an automatic
  categorization system for open source repositories, Journal of Systems and
  Software 79~(7) (2006) 939--953.
\newblock \href {http://dx.doi.org/10.1016/j.jss.2005.06.044}
  {\path{doi:10.1016/j.jss.2005.06.044}}.

\bibitem{Thummalapenta:2007:ParserWeb}
S.~Thummalapenta, T.~Xie, Parseweb: A programmer assistant for reusing open
  source code on the web, in: Proceedings of the IEEE/ACM Intl. Conf on ASE,
  ASE '07, ACM, New York, NY, USA, 2007, pp. 204--213.

\bibitem{2013SeidlVamos}
C.~Seidl, U.~Assmann, Towards modeling and analyzing variability in evolving
  software ecosystems, in: Proc. of the Int. Workshop on Variability Modelling
  of Software-intensive Systems (VaMoS '13), 2013.

\bibitem{Nonaka09SIGSEe}
M.~Nonaka, K.~Sakuraba, K.~Funakoshi, A preliminary analysis on corrective
  maintenance for an embedded software product family, IPSJ SIG Technical
  Report 2009-SE-166~(13) (2009) 1--8.

\bibitem{KimFSE2005}
M.~Kim, V.~Sazawal, D.~Notkin, G.~Murphy, An empirical study of code clone
  genealogies, in: Proceedings of the 10th European Software Engineering
  Conference Held Jointly with 13th International Symposium on Foundations of
  Software Engineering, 2005, pp. 187--196.

\end{thebibliography}


\begin{thebibliography}{2}
\expandafter\ifx\csname natexlab\endcsname\relax\def\natexlab#1{#1}\fi
\expandafter\ifx\csname url\endcsname\relax
  \def\url#1{\texttt{#1}}\fi
\expandafter\ifx\csname urlprefix\endcsname\relax\def\urlprefix{URL }\fi

\bibitem[{Ebert(2008)}]{EbertOSS}
Ebert, C., 2008. Open source software in industry. In: IEEE Software. pp.
  52--53.

\bibitem[{Hainemann et~al.(2011)Hainemann, Deissenboeck, Gleirscher, Hummel,
  and Irlbeck}]{HainemannICSR2011}
Hainemann, L., Deissenboeck, F., Gleirscher, M., Hummel, B., Irlbeck, M., 2011.
  On the extent and nature of software reuse in open source java projects. In:
  Proceedings of the 12th International Conference on Top Productivity Through
  Software Reuse. pp. 207--222.

\end{thebibliography}
 \end{document}